\documentclass[12pt,letterpaper]{article}
\usepackage{jheppub}
\pdfoutput=1
\usepackage{bbm}
\usepackage{amsmath, amsthm, amsfonts, amssymb}
\usepackage{mathrsfs} % for calligraphic C
\usepackage{dsfont}
\usepackage{wrapfig}
\usepackage{minibox}
\usepackage{graphicx}
\usepackage{xcolor}

\newcommand\Tr            {\mathrm{Tr}}
\newcommand\be            {\begin{equation}}
\newcommand\ee            {\end{equation}}
\newcommand\scr           {\scriptstyle}
\newcommand\ren           {R\'enyi\ }
\newcommand{\mt}[1]       {\textrm{\tiny #1}}

\def\({\left(} \def\){\right)}

\title{Charged \ren Entropies in CFTs with Einstein-Gauss-Bonnet Holographic Duals}
\author[a]{Georgios Pastras}
\author[b]{Dimitrios Manolopoulos}
\affiliation[a]{Department of Physics, School of Applied Mathematics and Physical Sciences, National Technical University, Athens 15780, Greece}
\affiliation[b]{Department of Nuclear and Particle Physics, Faculty of Physics, University of Athens, Athens 15784, Greece}
\emailAdd{georgios.pastras.physics@gmail.com}
\emailAdd{dimitrios.manolopoulos.physics@gmail.com}

\abstract{We calculate the \ren entropy $S_q(\mu,\lambda)$, for spherical entangling surfaces in CFT's with Einstein-Gauss-Bonnet-Maxwell holographic duals. \ren entropies must obey some interesting inequalities by definition. However, for Gauss-Bonnet couplings $\lambda$, larger than a specific value, but still allowed by causality, we observe a violation of the inequality $\frac{\partial }{{\partial q}}\left( {\frac{{q - 1}}{q}S_q(\mu,\lambda)} \right) \ge 0$, which is related to the existence of negative entropy black holes, providing interesting restrictions in the bulk theory. Moreover, we find an interesting distinction of the behaviour of the analytic continuation of $S_q(\mu,\lambda)$ for imaginary chemical potential, between negative and non-negative $\lambda$.}

\arxivnumber{1404.1309}

\begin{document}

\thispagestyle{empty}

\maketitle

\setcounter{footnote}{0}

\def\thefootnote{\arabic{footnote}}

%%%-----------------------------------------------------------------------------------------------------------------------------------------------------------------
\section{Introduction}
\label{sec:intro}

\ren entropy \cite{Renyi:1961,Renyi:1965} is used as a measure of entanglement for various quantum systems, ranging from condensed matter physics \cite{Levin:2006zz,Kitaev:2005dm,Hamma:2005,Calabrese:2004eu,Calabrese:2005zw} to Quantum Gravity and Holography \cite{Ryu:2006bv,Ryu:2006ef,Nishioka:2009un,Takayanagi:2012kg,VanRaamsdonk:2009ar,VanRaamsdonk:2010pw,Bianchi:2012ev,Myers:2013lva,Balasubramanian:2013rqa}. Entanglement entropy (EE) is also considered within the categorical approach to quantum mechanics and quantum computing (initiated by Abramsky and Coecke), where Frobenius algebras in a bicategory encode copying and deleting of classical information in quantum systems \cite{Abramsky:2003,Abramsky:2004,Abramsky:2005,Abramsky:2008,Coecke:2010,Baez:2011a,Vicary:2012,Baez:2014}. Interestingly, the quantum computation structures discussed in \cite{Vicary:2012} are formally nearly (exactly) those that were independently discussed in \cite{Carqueville:2012}, in the context of Topological Field Theories with defects.

Entanglement entropy in holographic theories first appear in \cite{Ryu:2006ef}, where an approach to calculate entanglement entropy of the boundary field theory was proposed. In this approach, the entanglement entropy between a spatial region $A$ and its complement in the $d$-dimensional boundary theory is proportional to the area of the $(d-1)$-dimensional minimal surface in the bulk that is homologous to $A$. This expression presents an interesting similarity with the usual expression for the thermal entropy of a black hole and although this surface does not necessarily coincide with an event horizon, there is much evidence supporting this conjecture \cite{Ryu:2006ef,Headrick:2010}.

It is important to pause for a while and say a few things about the task of computing entanglement entropy. In QFTs the standard approach involves the application of the so called \emph{Replica Trick}. Only a few simple (but not trivial) examples are known. In \cite{Calabrese:2004eu,Calabrese:2005zw} for example, they discuss the case of a finite interval of length $\ell$ in an infinite system and extend it to many other cases: finite systems, finite temperatures and to an arbitrary number of disjoint intervals, in a 2-dimensional CFT.

However, if we apply this construction for the boundary CFT in the holographic framework, we would end up with a conical singularity in the bulk with an angular excess of $2\pi(q-1)$, see for example the attempt in \cite{Fursaev:2006} and an explanation why it fails \cite{Headrick:2010}. Thus, the replica trick does not seem very helpful, without one's full understanding of string theory or quantum gravity in the AdS bulk, in deriving a holographic entanglement entropy formula. An alternative root was provided in \cite{Casini:2011kv} where they used conformal transformations to relate the entanglement entropy across a spherical entangling surface to the thermal entropy in a hyperbolic cylinder $\mathds{R}\times\mathds{H}^{d-1}$. We also follow this approach generalized for the grand canonical ensemble as in \cite{Caputa:2013,Belin:2013uta}.

The paper is organized as follows: In section \ref{sec:renyi}, we introduce \ren entropy for quantum systems. Interesting limits of \ren entropy are discussed and the generalization of \ren entropy for the grand canonical ensemble for a real and an imaginary chemical potential is given following \cite{Caputa:2013,Belin:2013uta}. In section \ref{sec:CFT}, we review the replica trick following \cite{Calabrese:2004eu,Calabrese:2005zw}, which allows the calculation of \ren entropy in CFT for 2-dimensional and higher dimensional theories. We then calculate the charged \ren entropy of a spherical entangling surface $S^{d-2}$ by the insertion of a Wilson loop encircling $S^{d-2}$ and in the limit $q \to 1$ we recover the thermal entropy. We also find formal expressions for the conformal dimension $h_q(\mu)$ and the magnetic response $k_q(\mu)$ of the generalized twist operators $\sigma_q(\mu)$, following \cite{Belin:2013uta}. In section \ref{sec:holographic}, the AdS/CFT correspondence \cite{Maldacena:1998,Witten:1998,Gubser:1998} is used to calculate the charged \ren entropy for CFTs with Einstein-Gauss-Bonnet gravity duals coupled to an electromagnetic field. The results are plotted for many interesting cases and a discussion follows about whether they are consistent with the Ryu-Takayanagi formula \cite{Ryu:2006bv,Ryu:2006ef}. We also see a violation of the inequality \eqref{eq:ren inequalities} which corresponds to a negative entropy black hole.

%%%-----------------------------------------------------------------------------------------------------------------------------------------------------------------
\section{Entanglement \ren Entropy}
\label{sec:renyi}

In this section, we review some aspects of \ren entropy that will be of interest to us in this paper. In particular, in subsection \ref{subsec:entanglement} we introduce some preliminary mathematical notions that will be of interest on later sections, while in subsection \ref{subsec:renyi_basics} we introduce the \ren entropy $S_q$ and its properties. In subsection \ref{subsec:renyi_grand_canonical} we generalize the definition of \ren entropy in the grand canonical ensemble according to the work of \cite{Caputa:2013,Belin:2013uta} to include a chemical potential $\mu$. The dependence of these charged \ren entropies $S_q(\mu)$ on the chemical potential, encodes the dependence of the entanglement on the charge.

%%%-----------------------------------------------------------------------------------------------------------------------------------------------------------------
\subsection{Entropy and Entanglement}
\label{subsec:entanglement}

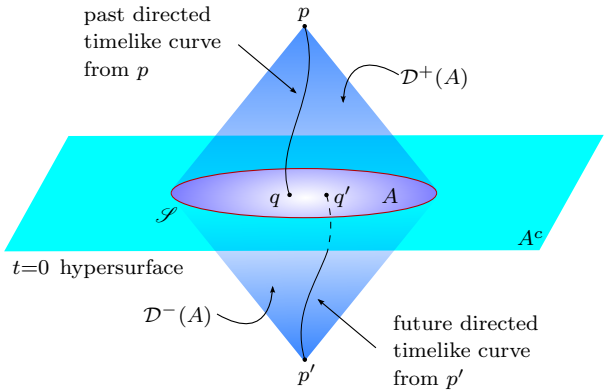
\begin{figure}[ht!]
\[
\raisebox{-58pt}{
  \begin{picture}(50,140)
    \put(-90,3){\scalebox{4}{\includegraphics{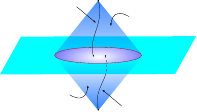}}}
  \put(-49,7){
    \setlength{\unitlength}{.90pt}\put(-32,-1){
    \put(153,115) {$\scr \mathcal{D}^+(A)$}
    \put(45,15) {$\scr \mathcal{D}^-(A)$}
    \put(110,143) {$\scr p$}
    \put(98,65) {$\scr q$}
    \put(125,65) {$\scr q'$}
    \put(145,65) {$\scr A$}
    \put(50,57) {$\scr \mathscr{S}$}
    \put(110,-10) {$\scr p'$}
    \put(202,47) {$\scr A^c$}
    \put(-10,35) {$\scr t=0$ \scriptsize{hypersurface}}
    \put(150,0) {\scriptsize{\minibox{future directed \\ timelike curve \\ from $p'$}}}
    \put(20,130) {\scriptsize{\minibox{past directed \\ timelike curve \\ from $p$}}}
         }\setlength{\unitlength}{1pt}}
  \end{picture}}
\]
\caption{\small{The division of the $t=0$ time slice into two regions $A$ and $A^c$ by the entangling surface $\mathscr{S}$. The Cauchy development $\mathcal{D}(A)=\mathcal{D}^+(A)\cup \mathcal{D}^-(A)$ is also shown.}}\label{fig:Cauchy development}
\end{figure}

Consider a quantum system in $d$-dimensional Minkowski spacetime $\mathds{R}^{1,d-1}$, composed by two subsystems, $A$ and its complement $A^c$. The space of states of the overall system is the tensor product $\mathcal{H}_\mt{A}\otimes_{\mathbb{C}}\mathcal{H}_{\mt{A}^c}$ of the respective Hilbert spaces of each subsystem. A mixed state is described by a density matrix $\rho$, i.e. an element in the Banach space $T(\mathcal{H}_\mt{A}\otimes_{\mathbb{C}}\mathcal{H}_{\mt{A}^c})$ of trace-class operators with unit trace, on the tensor product $\mathcal{H}_\mt{A}\otimes_{\mathbb{C}}\mathcal{H}_{\mt{A}^c}$. This overall quantum system is separated by an entangling surface $\mathscr{S}$ of codimension 2 at time $t=0$, as in figure \ref{fig:Cauchy development}, and suppose that this surface is closed and it is equipped with a metric $\gamma$. Then the system $A$ is described by the density matrix $\rho_\mt{A}\in T(\mathcal{H}_\mt{A})$, which can be found from the overall density matrix $\rho$ by tracing out the degrees of freedom of $A^c$, that is
\begin{equation}
\rho _\mt{A}  = {\mathop{\rm Tr}\nolimits} _{\mathcal{H}_{\mt{A}^c}} \rho ,
\end{equation}
where $\Tr_{\mathcal{H}_{\mt{A}^c}}$ denotes the \emph{partial trace} over $\mathcal{H}_{\mt{A}^c}$. It is easy to check that the partial trace viewed as a map, $\Tr_{\mathcal{H}_{\mt{A}^c}}\colon T(\mathcal{H}_\mt{A}\otimes_{\mathbb{C}}\mathcal{H}_{\mt{A}^c}) \to T(\mathcal{H}_\mt{A})$ is positive and trace preserving. Note also that $\Tr\rho_\mt{A}=\Tr\left( \Tr_{\mathcal{H}_{\mt{A}^c}} \rho \right)= \Tr \rho$ \cite[Sec.\,VI.6]{Reed:1980book}.

If the global system is defined in a curved space-time, the reduced density matrix is a functional of the metric $g$ of the space-time in the vicinity of $\mathscr{S}$, the intrinsic metric $\gamma$ and the extrinsic curvature $k$ of $\mathscr{S}$, that is $\rho_\mt{A}\equiv\rho_\mt{A}[g,\gamma,k]$ \cite{Solodukhin:2012zr}.

The future Cauchy development $\mathcal{D}^+(A)$ of $A$ is defined as the set of all points $p\in\mathds{R}^{1,d-1}$ such that every past-inextendible non-spacelike curve through $p$ intersects $A$ at least once. Similarly for the past Cauchy development $\mathcal{D}^-(A)$, see figure \ref{fig:Cauchy development}. The Cauchy development $\mathcal{D}(A)$ is the union of the future and past Cauchy developments. Knowledge of the appropriate data on the closed set $A$ (if one knew data on an open set, that in its closure would follow from continuity) would determine events in $\mathcal{D}^+(A)$ \cite[Sec.\,6.5]{Hawking:1973book}.

The density matrix of $A^c$ may describe a mixed state, even if the overall density matrix $\rho$ describes a pure state. This occurs because of the possible \emph{entanglement} between the two subsystems. Thus, the spectrum of the density matrix $\rho_\mt{A}$ contains information about the aforementioned entanglement. Let's implement a trivial clarifying example.

Suppose that both $A$ and $A^c$ are a single spinor. If the whole system is described by the non-entangled pure state $\left|\psi\right\rangle = \frac{1}{2}\left({\left({\left|\uparrow\right\rangle + \left|\downarrow\right\rangle}\right)_\mt{A}}\otimes{\left({\left|\uparrow\right\rangle + \left|\downarrow\right\rangle}\right)_{\mt{A}^c}}\right)$, then the density matrix of system $A$ is
\be\label{eq:example1}
\rho_\mt{A} = \frac{1}{2}\left({\left|\uparrow\right\rangle + \left|\downarrow\right\rangle }\right)\left({\left\langle\uparrow\right| + \left\langle\downarrow\right|}\right),
\ee
which is also a pure state. Thus, in this case the eigenvalues of $\rho _\mt{A}$ are 0 and 1.

On the other hand, if the whole system is described by the maximally entangled pure state $\left| \psi  \right\rangle  = \frac{1}{{\sqrt 2 }}\left( {{{\left|  \uparrow  \right\rangle }_\mt{A}} \otimes {{\left|  \uparrow  \right\rangle }_{\mt{A}^c}} + {{\left|  \downarrow  \right\rangle }_\mt{A}} \otimes {{\left|  \downarrow  \right\rangle }_{\mt{A}^c}}} \right)$, then the density matrix of system $A$ is
\begin{equation}\label{eq:example2}
{\rho _\mt{A}} = \frac{1}{2}\left( {\left|  \uparrow  \right\rangle \left\langle  \uparrow  \right| + \left|  \downarrow  \right\rangle \left\langle  \downarrow  \right|} \right),
\end{equation}
which is a mixed state with the density matrix having two eigenvalues equal to $\frac{1}{2}$. However, one should notice that the same density matrix for system $A$ could have occurred if the overall system was in a non-entangled mixed state, for example,
\begin{equation}\label{eq:example3}
\rho  = \frac{1}{2}\left( {{{\left|  \uparrow  \right\rangle }_\mt{A}} \otimes {{\left|  \uparrow  \right\rangle }_{\mt{A}^c}}{{\left\langle  \uparrow  \right|}_\mt{A}} \otimes {{\left\langle  \uparrow  \right|}_{\mt{A}^c}} + {{\left|  \downarrow  \right\rangle }_\mt{A}} \otimes {{\left|  \downarrow  \right\rangle }_{\mt{A}^c}}{{\left\langle  \downarrow  \right|}_\mt{A}} \otimes {{\left\langle  \downarrow  \right|}_{\mt{A}^c}}} \right).
\end{equation}

The above example suggests that, assuming the overall system is in a pure state, the entanglement between subsystems $A$ and $A^c$ can be described by the spectrum of $\rho_\mt{A}$ and specifically the higher the entanglement between the two subsystems, the more ``disordered'' the spectrum of $\rho_\mt{A}$ appears to be. As in statistical mechanics, a logarithmic measure of the ``disorder'' is the Shannon entropy, usually called von Neummann entropy when applied to the spectrum of a density matrix.

The entanglement entropy is defined as the von Neumann entropy corresponding to the density matrix $\rho_\mt{A}$,
\begin{equation}
S_\mt{EE}  :=  - {\mathop{\rm Tr}\nolimits} \left( {\rho_\mt{A} \ln \rho _\mt{A} } \right) .
\label{eq:entanglement_entropy_definition}
\end{equation}

\subsection{Entanglement \ren Entropy}
\label{subsec:renyi_basics}

A problem that usually appears is that one does not have a good way to represent the operator $\ln \rho_\mt{A}$ and thus to calculate entanglement entropy. However, if one represents the density matrix of the full system by a path integral (as in the vacuum or a thermal ensemble, for example), then the reduced density matrix $\rho_\mt{A}$ and its positive integer powers $\rho_\mt{A}^q$ can also be represented by path integrals. If those path integrals can be computed explicitly for all $q$, then one can obtain a generalization of Entanglement entropy called Entanglement \ren entropy (or just \ren entropy from now on). The entanglement entropy can then be indirectly calculated as the limit $q\to 1$ of the \ren entropy.

\ren entropies are defined as the moments of the density matrix $\rho_\mt{A}$ which describes system $A$, as
\begin{equation}
S_q := \frac{1}{{1 - q}}\ln {\mathop{\rm Tr}\nolimits} \rho_\mt{A} ^q , \ \ q\in\mathds{R}^+ -\{1\}.
\label{eq:Renyi_entropy_definition}
\end{equation}
One can recover the entanglement entropy from the \ren entropies as an appropriate limit
\begin{equation}
S_1 \equiv \mathop {\lim }\limits_{q \to 1} S_q = S_\mt{EE} .
\end{equation}
\ren entropy contains also some other interesting limits. Specifically, the limit $q \to 0$ recovers the Hartley entropy
\begin{equation}
 S_0 \equiv \mathop {\lim }\limits_{q \to 0} S_q  = \ln D ,
 \label{eq:hartley}
 \end{equation}
where $D$ is the number of non-vanishing eigenvalues of the density matrix $\rho_\mt{A}$. The limit $q \to \infty$ recovers the Min-entropy
\begin{equation}
S_{\infty} \equiv \mathop {\lim }\limits_{q \to \infty } S_q  =  - \ln \lambda  ,
\label{eq:min_entropy}
 \end{equation}
where $\lambda\in \mathbb{C}$ is the greatest eigenvalue of $\rho_\mt{A}$.
An interesting specific value of \ren entropy is also the $q = 2$ case, called the Collision entropy
\begin{equation}
S_2  \equiv -\ln {\mathop{\rm Tr}\nolimits} \rho_\mt{A} ^2 = - \ln P ,
\end{equation}
where $P$ is the probability to find two systems described by density matrix $\rho_\mt{A}$ in the same state after measurements in the basis that diagonalizes $\rho_\mt{A}$.

\ren entropies, apart from providing an indirect method to calculate entanglement entropy, are interesting for other reasons. It can be shown that the knowledge of all \ren entropies for all integers $q>0$ is sufficient to recover the whole spectrum of the density matrix $\rho_\mt{A}$ \cite{Calabrese:2008es}. Thus, \ren entropies contain much more information about entanglement than entanglement entropy.

\ren entropies obey the following four interesting inequalities.
\begin{align}
\frac{{\partial {S_q}}}{{\partial q}} &\le 0,\\
\frac{\partial }{{\partial q}}\left( {\frac{{q - 1}}{q}{S_q}} \right) &\ge 0, \label{eq:ren inequalities}\\
\frac{\partial }{{\partial q}}\left( {\left( {q - 1} \right){S_q}} \right) &\ge 0,\\
\frac{{{\partial ^2}}}{{\partial {q^2}}}\left( {\left( {q - 1} \right){S_q}} \right) &\le 0.
\end{align}
These inequalities are a direct consequence of \ren entropy definition:
\begin{equation}
{S_q} := \frac{1}{{1 - q}}\ln \sum_{i\in \mathcal{I}} {p_i^q} \ \ \mathrm{with} \ \ 0 \le {p_i} \le 1 \ \ \mathrm{and} \ \ \sum_{i \in \mathcal{I}} {p_i}=1.
\end{equation}
When we proceed to calculate \ren entropies of CFTs from holographic dual theories, these inequalities can provide interesting constraints in the former, as in this case, the calculation is not performed directly on a probability distribution, but rather based on black hole thermodynamics, making the validity of the above inequalities non-trivial.

%%%-----------------------------------------------------------------------------------------------------------------------------------------------------------------
\subsection{\ren Entropy in the Grand Canonical Ensemble}
\label{subsec:renyi_grand_canonical}

In this paper, theories in the grand canonical ensemble with a conserved charge are considered. The definition of \ren entropy requires an extension as described in \cite{Caputa:2013,Belin:2013uta}, that is
\begin{equation}\label{eq:charged_Renyi_entropy_definition}
S_q (\mu) := \frac{1}{{1 - q}}\ln \Tr\varrho_\mt{A}^q(\mu) , \ \ \ \varrho_\mt{A}(\mu)\equiv{\rho _\mt{A} \frac{{e^{\mu Q_\mt{A} } }}{{n_\mt{A} \left( \mu  \right)}}},
\end{equation}
where $n_\mt{A}(\mu)\equiv \Tr\left[ {\rho _\mt{A} e^{\mu Q_\mt{A} } } \right]$ ensures that the introduced grand canonical density matrix $\varrho_\mt{A}(\mu)$ is appropriately normalized having unit trace. One can also define charged \ren entropies for an imaginary chemical potential
\begin{equation}\label{eq:charged_Renyi_entropy_definition_imaginary}
\tilde{S}_q (\mu_\mt{E}) := \frac{1}{{1 - q}}\ln \Tr\tilde{\varrho}_\mt{A}^q(\mu_\mt{E}) , \ \ \ \tilde{\varrho}_\mt{A}(\mu_\mt{E})\equiv\varrho_\mt{A}(i\mu_\mt{E}) ,
\end{equation}
with $\mu_\mt{E}\in\mathds{R}$. When we analytically continue from \eqref{eq:charged_Renyi_entropy_definition} to \eqref{eq:charged_Renyi_entropy_definition_imaginary}, typically, only some singularities appear in the imaginary $\mu$-axis.

Studying imaginary chemical potential can provide information about the confinement phase transition in QCD theories with fermions \cite{Roberge:1986,de Forcrand:2002ci,de Forcrand:2003hx}. In such theories, the related $U(1)$ symmetry is the one that corresponds to the fermion number conservation. Then, if one considers the partition function with an imaginary potential, they find
\begin{equation}
\tilde{Z}\left( {\mu_\mt{E}} \right) = \Tr\left[e^{ - \frac{E}{{{k_b}T}} + i\mu_\mt{E}B}\right] ,
\end{equation}
where $B = \int \! {{d^3}x \ {\psi ^\dag }\psi } $ is the baryon number. Due to the fact that the baryon number is quantized, it is obvious that $\tilde{Z}\left( \mu_\mt{E} \right)$ is a periodic function of $\mu_\mt{E}$ with period $2 \pi$. However, this is true only in the deconfined phase of QCD. In the confining phase of QCD, the fermionic states are not allowed to have color charge resulting in the formation of bound states. If an $SU(N)$ QCD is considered, then these bound states in the \emph{deconfined phase} are characterized by baryon number $B \in \mathds{N}_0$ while in the \emph{confined phase} by $B\in N\mathds{N}_0$, for example mesons have zero baryon number, while baryons like protons or neutrons have baryon number equal to $N$. The above mean that in the confining phase of QCD, $\tilde{Z}\left( \mu_\mt{E} \right)$ is a periodic function of $\mu_\mt{E}$ with period $\frac{2 \pi}{N}$. Moreover, the imaginary chemical potential can be used to avoid the sign problem in lattice algorithms \cite{Alford:1998sd,D'Elia:2002gd,D'Elia:2004at}.

%%%-----------------------------------------------------------------------------------------------------------------------------------------------------------------
\section{CFT Computation of \ren Entropy}
\label{sec:CFT}

In this section, we will review in some detail the QFT approach to entanglement entropy following \cite{Calabrese:2004eu,Calabrese:2005zw}. In particular, we start in subsection \ref{subsec:replica 2d} by reviewing the \emph{replica trick} in a 2-dimensional CFT. In short, the replica trick provides a useful way to represent the density matrix $\rho$ of a quantum system, which is in a state (such as the vacuum or a thermal ensemble), as a path integral over a Euclidean spacetime $\mathds{R}^2$. Then, the reduced density matrix $\rho_\mt{A}$ and its positive integer powers $\rho_\mt{A}^q$ can be represented as path integrals. Recall that the \ren entropy is defined as $S_q :=\frac{\ln\rho_\mt{A}^q}{1-q}$, for $q>1$. If we analytically continue for $q$ and we take the limit $q\to 1$, we obtain in this way, indirectly, the entanglement entropy. This is done in subsection \ref{subsec:correnators of twist operators} following \cite{Calabrese:2004eu,Calabrese:2005zw} by virtue of the so called twist operators $\sigma_q$. In subsection \ref{subsec:Charged Ren Entropy}, following \cite{Belin:2013uta,Casini:2011kv,Hung:2011nu}, we discuss charged \ren entropies for spherical entangling surfaces in $d$-CFT and we derive a formula for the charged \ren entropy $S_q(\mu)$ in terms of the thermal entropy $S(T)$ on a hyperbolic space $S^1\times\mathds{H}^{d-1}$. Finally, in subsection \ref{subsec:Correlators of Generalized Twist Operators}, we briefly review the notion of generalized conformal dimension and magnetic response of the underlying generalized twist operators $\sigma_q(\mu)$.

%%%-----------------------------------------------------------------------------------------------------------------------------------------------------------------
\subsection{Replica Trick in 2\emph{d}-CFT}
\label{subsec:replica 2d}

We now turn our attention to two dimensional CFTs. These are Euclidean QFTs whose symmetry group contains, in addition to the Euclidean symmetries, local conformal transformations, i.e. transformations that preserve angles but not necessarily lengths. Indeed, in two dimensions there exists an infinite variety of coordinate transformations that, although not everywhere well defined, are locally conformal and they are holomorphic mappings $z\mapsto w(z)=z+\epsilon(z)$, from the complex plane to itself. The local conformal symmetry is of special importance in two dimensions since the corresponding symmetry algebra is infinite-dimensional and in principle, one can calculate everything in an analytic way.

\begin{figure}[ht!]
\[
\raisebox{-40pt}{
  \begin{picture}(100,100)
  \put(-60,3){\scalebox{2.5}{\includegraphics{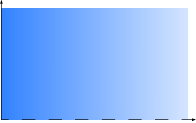}}}
  \put(0,-10){
    \setlength{\unitlength}{.90pt}\put(-32,-1){
    \put(-42,7) {0}
    \put(40,23) {$\scr\phi(z)$}
    \put(40,97) {$\scr\phi'(z')$}
    \put(-16,7) {$\scr v_1$}
    \put(15,7) {$\scr \cdots$}
    \put(49,7) {$\scr v_i$}
    \put(70,7) {$\scr \cdots$}
    \put(91,7) {$\scr v_n$}
    %\put(20,60) {$z$ - plane}
    \put(120,7) {$x$}
    \put(-40,115) {$i\tau$}
    \put(-42,103) {$\scr \beta$}
         }\setlength{\unitlength}{1pt}}
  \end{picture}}
\xrightarrow{\scr \frac{\beta}{2 \pi i} \ln z}
\raisebox{-40pt}{
  \begin{picture}(120,100)
  \put(19,5){\scalebox{2.5}{\includegraphics{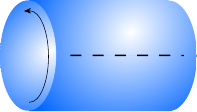}}}
  \put(0,-10){
    \setlength{\unitlength}{.90pt}\put(-32,-1){
    \put(146,68) {$\scr\phi(\zeta)$}
    \put(146,52) {$\scr\phi'(\zeta)$}
    \put(80,55) {$\tau$}
    \put(135,9) {$\longrightarrow$}
    \put(140,3) {$x$}
      }\setlength{\unitlength}{1pt}}
  \end{picture}}
\]
\caption{\small{The map from the complex plane to the infinite cylinder of circumference $\beta$, where $\rho_\mt{A}$ is represented as a path integral.}}\label{fig:Cplane to Cylinder map}
\end{figure}
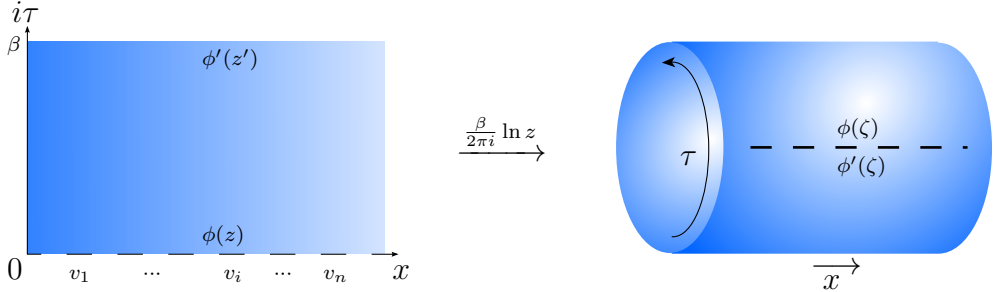

Consider now a two dimensional quantum system on the complex plane parameterized by $z=x+i\tau$. Let $A$ be the subsystem consisting of the union of disjoint intervals $v_i=[a_i,b_i]$, where $a_i < b_i < a_{i+1}$, along the real line $\tau = 0$, i.e. $A=\bigcup_{i=1}^n v_i$. In this case the entangling surface $\mathscr{S}$ consists of the end points of the disjoint intervals. The partition function of the system is $Z(\beta)=\Tr \ e^{-\beta H}$, where $\beta$ is the inverse temperature and $H$ is the Hamiltonian operator. An expression for the reduced density matrix $\rho_\mt{A}$ can be found by identifying along $\tau=0$ and $\tau=\beta$ and sewing together only the points of $A^c$, which will leave open cuts, one for each interval $v_i$ along the line $\tau=0$. This can be done via the map $z\mapsto \zeta(z)=\frac{\beta}{2 \pi i} \ln z$ from the complex plane parameterized by $z$, to the infinite cylinder of circumference $\beta$, parameterized by $\zeta$, see figure \ref{fig:Cplane to Cylinder map}, where now $x\in\mathds{R}$ runs along the flat direction of the cylinder and time is compactified to $\tau\in[0,\beta]$.\footnote{The Hamiltonian operator on the cylinder is now given as the sum of the zero modes of the holomorphic and anti-holomorphic component of the stress tensor, $H_{\mathrm{cyl}}= \frac{2 \pi}{\beta} \left(L_0 + \bar{L}_0 - \frac{c}{12} \right)$, where $c$ is the central charge of the CFT.} For any two field insertions $\phi(\zeta)$ and $\phi'(\zeta')$ this means that $\phi(\zeta)=\phi'(\zeta)$ on the branch cuts.

One can now compute $\Tr \rho_\mt{A}^q$ for any $q\in\mathds{N}$ by making $q$ copies of the above, labeled by $s\in\mathds{N}$ with $1\leq s \leq q$ and sewing them cyclically along the cuts so that $\phi_{s}(\zeta)=\phi'_{s+1}(\zeta)$ and $\phi_{q}(\zeta)=\phi'_{1}(\zeta)$ for all $\zeta \in A$. The various copies talk to each other only along the $q$-fold line segments introduced along $\tau = 0$.

%%%-----------------------------------------------------------------------------------------------------------------------------------------------------------------
\subsection{Correlators of Twist Operators}
\label{subsec:correnators of twist operators}

The replica trick discussed above, can be realized with the insertion of the so called twist operators $\sigma_{\pm q}(p_i)$ at the end-points $p_i$ of the interval $v_i$, in the path integral over the replicated theory. These operators are responsible for opening and closing the branch cuts at the end-points of the various intervals. In particular, in two dimensional CFT, the twist operators $\sigma_{q}(p_i)$ and $\sigma_{-q}(p_i)$ are (spinless) primary fields with the same scaling dimension $\Delta_q=\bar{\Delta}_q$.\footnote{In $d$ dimensions, the twist operators will be some $(d-2)$-dimensional \emph{surface operators}, in agreement with \cite{Ryu:2006bv}.} Then, the trace of the $q$-th power of the reduced density matrix can be written as a path integral
\be\label{eq:Tr rho path int}
\Tr \rho_\mt{A}^q = \frac{1}{Z_1^{q}}\int \! \prod_{s=1}^q \left[\mathcal{D}\phi_s(z) \right] \  \sigma_q(p_1)\sigma_{-q}(p_1)\ldots\sigma_q(p_n)\sigma_{-q}(p_n) e^{-\sum_{r=1}^q S_r[\phi(z)]} \equiv \frac{Z_q}{Z_1^q},
\ee
where $S_r=\int_0^\beta \! L_r \ d\tau$, with $L_r$ the Euclidean lagrangian on each copy of the $q$-sheeted structure, $Z_1^q$ is the vacuum functional of the original theory and $Z_q$ is the functional on the $q$-sheeted orbifold space which we will denote by $\mathcal{O}_q$. With this notation, equation \eqref{eq:Tr rho path int} is simply the correlator of the twist operators on $\mathcal{O}_q$.
\be\label{eq:Tr as corr of sigma}
\Tr \rho_\mt{A}^q  = \left\langle \sigma_q(p_1)\sigma_{-q}(p_1)\ldots\sigma_q(p_n)\sigma_{-q}(p_n) \right\rangle_{\mathcal{O}_q}.
\ee
This correlator is divergent because of the singular geometry on $\mathcal{O}_q$ at the branch points. However, in two dimensional CFT this is not a problem because every metric is conformally flat and therefore, the singular metric on $\mathcal{O}_q$ can be mapped to a flat metric via a conformal transformation. The transformation we need is the conformal mapping
\be\label{eq:mobius transformation}
\zeta\mapsto w(\zeta)=\prod_{i=1}^n\left(\frac{\zeta-a_i}{\zeta-b_i}\right)^{\frac{1}{q}},
\ee
which maps the singular $\zeta$-plane to the smooth $w$-plane. In particular, we have exploded the cylinder to the Riemann sphere $\mathbb{C}_{\infty}\equiv\mathbb{C}\cup\{\infty\}$ parameterized by $w$, where the zero point on $\mathbb{C}_{\infty}$ is given by $w(a_i) = 0$ and the point at infinity by $w(b_i)\to \infty$. The power $1/q$ smooths out the singularity in $\mathcal{O}_q$ by patching together $q$ copies of $\zeta$-planes. Observe also that $\lim_{\zeta\to\infty} w(\zeta)$ is mapped to the $q$-th root of unity in $\mathbb{C}_{\infty}$. Note that for $q=1$ and $n=1$ the mapping \eqref{eq:mobius transformation} is just a M\"{o}bius transformation.

The correlator of the twist operator in CFT is given by the operator product expansion (OPE) of the holomorphic part of the stress tensor $T(\zeta)$ with the twist operators (conformal Ward identity). The stress tensor $T(\zeta)$ under the conformal transformation \eqref{eq:mobius transformation} transforms as a \emph{quasi primary} field according to
\be\label{eq:transformation of T}
T(\zeta) = \left(\frac{dw}{d\zeta}\right)^{2}T(w) + \frac{c}{12}\left\{ w;\zeta \right\},
\ee
where $c$ is the \emph{central charge} of the CFT and $\left\{ w;\zeta \right\}:=\tfrac{w'''(\zeta)}{w'(\zeta)}-\tfrac{3}{2}\left( \tfrac{w''(\zeta)}{w'(\zeta)} \right)^2$, is the \emph{Schwarzian derivative}. By translational and rotational invariance, the expectation value of the stress tensor on $\mathbb{C}_{\infty}$ vanishes, i.e. $\left\langle T(w) \right\rangle_{\mathbb{C}_{\infty}}=0$, thus we are left to calculate only the Schwarzian derivative. In order to illustrate this, we do the calculation for a single interval $[a,b]$, which yields
\be\label{eq:<T(zeta)>}
\left\langle T(\zeta) \right\rangle_{\mathcal{O}_q}=\frac{cq}{24}\left( 1-\frac{1}{q^2}\right)\frac{(b-a)^2}{(\zeta-a)^2(\zeta-b)^2}.
\ee
Note that in calculating the Schwarzian derivative, we get the above expression without the overall factor of $q$. This is because $T(\zeta)$ is inserted on a single sheet of the $q$-sheeted orbifold structure. If we insert $T(\zeta)$ on all the sheets, one obtains \eqref{eq:<T(zeta)>}.

We now compute the expectation value of the OPE of the stress tensor with the two primary fields $\sigma_q(a)$ and $\sigma_{-q}(b)$ which have the same holomorphic and anti-holomorphic scaling dimension
\be\label{eq:scaling dim}
\Delta_q=\bar{\Delta}_q=\frac{c}{24}\left( 1-\frac{1}{q^2} \right).
\ee
We assume that $\zeta$ is a complex coordinate on a single sheet $\mathbb{C}_{\infty}$, which is now decoupled from the others. Then, the conformal Ward identity reads
\be\label{eq:<T(zeta)ss>}
\left\langle T(\zeta)\sigma_q(a)\sigma_{-q}(b) \right\rangle_{\mathbb{C}_{\infty}}=\frac{h_q/2}{(\zeta-a)^2(\zeta-b)^2(b-a)^{2h_q-2}}, \ \ \mathrm{where} \ \ h_q\equiv \Delta_q+\bar{\Delta}_q .
\ee
This result is for a single sheet, inserting thus $T(\zeta)$ on all the sheets, the right hand side of \eqref{eq:<T(zeta)ss>} is multiplied by a factor of $q$. Furthermore, the twist operators are normalized so that their two point function is $\left\langle \sigma_q(a)\sigma_{-q}(b) \right\rangle_{\mathbb{C}_{\infty}}=|b-a|^{-2h_q}$. Comparing equations \eqref{eq:<T(zeta)>} and \eqref{eq:<T(zeta)ss>} we see that
\be\label{eq:<T(zeta)>equality}
\left\langle T(\zeta) \right\rangle_{\mathcal{O}_q}=\frac{\left\langle T(\zeta)\sigma_q(a)\sigma_{-q}(b) \right\rangle_{\mathbb{C}_{\infty}}}{\left\langle\sigma_q(a)\sigma_{-q}(b) \right\rangle_{\mathbb{C}_{\infty}}}.
\ee
Under conformal transformations $\zeta \mapsto \xi(\zeta)$, the Ward identity \eqref{eq:<T(zeta)ss>} determines all the properties of the correlation function. Therefore, we conclude that under scale and conformal transformations $\Tr \rho_\mt{A}^q$ behaves exactly as the $q$-th power of the two point function of the twist fields, that is
\be\label{eq:2-point qth power}
\Tr \rho_\mt{A}^q = \left\langle \sigma_q(a)\sigma_{-q}(b) \right\rangle_{\mathcal{O}_q}=C_q\left(\left\langle \sigma_q(a)\sigma_{-q}(b) \right\rangle_{\mathbb{C}_\infty}\right)^q,
\ee
up to an overall constant $C_q$, which is not determined by conformal symmetry alone but from the details of the CFT. However, since $\Tr \rho_\mt{A}^q=1$ we see that $C_1=1$. Therefore, we have that
\be\label{eq:Tr qth power}
\Tr \rho_\mt{A}^q = C_q\left( \frac{b-a}{\delta} \right)^{-2qh_q},
\ee
where $\delta$ is the short-distance UV cut-off. Therefore, the \ren entropy \eqref{eq:Renyi_entropy_definition} becomes
\be\label{eq:Renyi entropy from replica trick}
S_q = \frac{c}{6}\left( 1+ \frac{1}{q} \right) \ln\frac{\ell}{\delta} + \frac{\ln C_q}{1-q}, \ \ \mathrm{where} \ \ \ell\equiv b-a.
\ee
Taking the limit $q\to 1$, we recover the entanglement entropy
\be\label{eq:q to 1 EE}
\lim_{q\to1} S_q =  \frac{c}{3}\ln\frac{\ell}{\delta} - \left.\frac{dC_q}{dq}\right|_{q=1} = S_\mt{EE}.
\ee

%%%-----------------------------------------------------------------------------------------------------------------------------------------------------------------
\subsection{Charged \ren Entropy for Spherical Entangling Surfaces}
\label{subsec:Charged Ren Entropy}

In this section, following \cite{Belin:2013uta,Casini:2011kv}, we will briefly review some aspects of charged \ren entropies for a quantum system in $d$-dimensional Euclidian space-time $\mathds{R}^d$. This quantum system consists of two subsystems $A$ and $A^c$, as before, which are separated by a spherical entangling surface $\mathscr{S}=S^{d-2}$.

In \cite{Casini:2011kv}, it was shown that the entanglement entropy is related to the thermal entropy $S(T)$ of the CFT on the hyperbolic cylinder $\mathds{R}\times\mathds{H}^{d-1}$, with the temperature and curvature fixed by the radius of curvature $R$ of $S^{d-2}$. This was done by conformally mapping the Cauchy development\footnote{In \cite{Casini:2011kv} it is referred to as ``causal development".} $\mathcal{D}(A)$ of the region $A$ on the inside of $S^{d-2}$ to $\mathds{R}\times\mathds{H}^{d-1}$. In this way, the vacuum correlators in $\mathcal{D}(A)$ are conformally mapped to thermal correlators in $\mathds{R}\times\mathds{H}^{d-1}$. Under this conformal transformation, the density matrix which describes the CFT vacuum state inside $S^{d-2}$ is mapped to a thermal bath in $\mathds{R}\times\mathds{H}^{d-1}$.

To see this, we perform the conformal transformation which maps $\mathcal{D}(A)$ to the Euclidian background $S^1\times\mathds{H}^{d-1}$ instead, where the time coordinate $t_\mt{E}$ is compactified on $S^1$. We start with the flat space metric in polar coordinates:
\be\label{eq:metric R^d in polar form}
ds^2_{\mathds{R}^d}= dt_\mt{E}^2 + dr^2 + r^2dS_{d-2}^2,
\ee
where $r$ is the radial coordinate on the constant time slices and $dS_{d-2}^2$ is the metric on $S^{d-2}$. The spherical entangling surface is at $(t_\mt{E},r)=(0,R)$. We can rewrite this in a complex form by using a complex coordinate $z = r+it_\mt{E}$, then
\be\label{eq:metric R^d in z form}
ds^2_{\mathds{R}^d}= dz d\bar{z} + \( \frac{z+\bar{z}}{2} \)^2dS_{d-2}^2.
\ee
Making the following coordinate transformations
\be\label{eq: z to w transformation}
e^{-w}=\frac{R-z}{R+z} \ \ \mathrm{and} \ \ w = u + i\frac{\tau_\mt{E}}{R},
\ee
the above metric becomes
\be\label{eq:metric R^d-1 to H^d-1}
ds^2_{S^1\times\mathds{H}^{d-1}}= \Omega^2 ds^2_{\mathds{R}^d} = d\tau_\mt{E}^2 + R^2\(du^2 + \sinh^2u \ dS_{d-2}^2\),
\ee
where $\Omega$ is the conformal factor
\be\label{eq:conf mapping from D to H}
\Omega = \frac{2R^2}{\left| R^2 - z^2 \right|} = \left| 1 + \cosh w \right|.
\ee
Hence, the mapping \eqref{eq:conf mapping from D to H} is a conformal mapping from $\mathcal{D}(A)$ to $S^1\times\mathds{H}^{d-1}$. In particular, it takes the CFT in the Minkowski vacuum on $\mathcal{D}(A)$ to a thermal state on $\mathds{R}\times\mathds{H}^{d-1}$, with a physical temperature
\be\label{eq:T_0}
\beta_0^{-1} \equiv T_0 = \frac{1}{2 \pi R}.
\ee
The thermal density matrix in the new spacetime $S^1\times\mathds{H}^{d-1}$ is related to that on $\mathcal{D}(A)$ by a unitary transformation $U$. More explicitly, we may write the density matrix on $\mathcal{D}(A)$ as
\be\label{eq:unitary transform or rho}
\rho_\mt{A} \mapsto U^{-1}\frac{e^{-\beta_0 H}}{Z(\beta_0)}U, \ \ \mathrm{where} \ \ Z(\beta) \equiv \Tr \ e^{-\beta H}.
\ee
Note that in taking the trace of equation \eqref{eq:unitary transform or rho}, the unitary transformation $U$ and its inverse cancel out. Thus, the entanglement entropy across the sphere becomes the thermal entropy $S(T)$ in $\mathds{R}\times\mathds{H}^{d-1}$. The idea now is that if one varies the temperature, then the thermal entropy calculates \ren entropies \cite{Hung:2011nu,Klebanov:2011uf}.

When considering charged \ren entropies as in \cite{Caputa:2013,Belin:2013uta} for example, the current operator $J_\mu$ in the underlying CFT is also considered. This operator is associated with the conserved global charge
\be\label{eq:global charge}
Q_\mt{A} = \int_\mt{A} \! d^{d-1}x \ J_0.
\ee
This global symmetry in the boundary CFT, according to \cite{Belin:2013uta}, gives rise to a gauge field in the dual gravity theory under consideration and thus, $S_q(\mu)$ is related to the thermal entropy of a charged hyperbolic black hole. In terms of the boundary CFT, this means that in order to extend the path integral calculations of $S_q$ to include a chemical potential, one needs to insert a Wilson loop encircling the entangling surface. The relevant fixed background gauge potential $B_\mu$, coupled to the conserved current $J_\mu$, represents the chemical potential. However, in the thermal path integral, the Euclidean time direction is compactified with period $\beta_0 \equiv T^{-1}_0$ and then the chemical potential appears by inserting a nontrivial Wilson line
\be\label{eq:wilson loop exp}
W(\mathcal{C}):= \exp \(i Q_\mt{A}\oint_{\mathcal{C}}\! dx^\mu \ B_\mu \) = e^{i\mu_\mt{E} Q_\mt{A}},
\ee
on this thermal circle. Note, that since we are discussing the abelian case, we do not need the trace and the exponential does not need to be path ordered.

To understand the physical meaning of the Wilson loop, consider the path integral description of a field theory with a field $\phi$ and an action $S[\phi]$ on $S^{d-2}$. Then, consider the general unperturbed correlation function on $S^{d-2}$
\be\label{eq:unperturbed sphere corr}
\left\langle
\raisebox{-27.5pt}{
  \begin{picture}(60,65)
  \put(-4,0){\scalebox{2}{\includegraphics{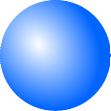}}}
  \put(-20,-7.5){
    \setlength{\unitlength}{.90pt}\put(-32,-1){
      }\setlength{\unitlength}{1pt}}
  \end{picture}} \right\rangle
= \frac{1}{Z}\int \! [\mathcal{D}\phi] \ \phi(x_1)\ldots\phi(x_n) \ e^{-iS[\phi]},
\ee
where $Z$ is the vacuum functional. One can now modify the action by terms localised on a line. For example, if $\phi$ lives on a circle with circumference $\beta$ on $S^{d-2}$, one could replace $iS[\phi]$ by $iS[\phi] + i Q_\mt{A}\int_{0}^\beta \! dx^\mu \ B_\mu $ for some gauge potential $B_\mu$. More explicitly, the correlation function in the perturbed mode would be
\be\label{eq:perturbed sphere corr}
\left\langle
\raisebox{-27.5pt}{
  \begin{picture}(60,65)
  \put(-4,0){\scalebox{2}{\includegraphics{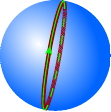}}}
  \put(-20,-7.5){
    \setlength{\unitlength}{.90pt}\put(-32,-1){
    \put(130,30) {pert}
             }\setlength{\unitlength}{1pt}}
  \end{picture}}
\right\rangle_{\phantom{\mathrm{pert}}} = \frac{1}{Z_{\mathrm{pert}}}\int \! [\mathcal{D}\phi] \ \phi(x_1)\ldots\phi(x_n) \ e^{-\(iS[\phi]+i Q_\mt{A}\int_{0}^\beta \! dx^\mu \ B_\mu \)}.
\ee
Observe now that
\be\label{eq:wilson loop corr}
\left\langle
\raisebox{-27.5pt}{
  \begin{picture}(60,65)
  \put(-4,0){\scalebox{2}{\includegraphics{perturbed_sphere}}}
  \put(-20,-7.5){
    \setlength{\unitlength}{.90pt}\put(-32,-1){
    \put(130,30) {pert}
             }\setlength{\unitlength}{1pt}}
  \end{picture}}
\right\rangle_{\phantom{\mathrm{pert}}}
=
\left\langle
\raisebox{-27.5pt}{
  \begin{picture}(60,65)
  \put(-4,0){\scalebox{2}{\includegraphics{sphere}}}
  \put(-20,-7.5){
    \setlength{\unitlength}{.90pt}\put(-32,-1){
      }\setlength{\unitlength}{1pt}}
  \end{picture}}
\ W(\mathcal{C})\right\rangle.
\ee
Thus the perturbed correlator is obtained by inserting the Wilson loop \eqref{eq:wilson loop exp} into the unperturbed one.

Therefore, the twist operators $\sigma_{\pm q}$ which appear in the replica trick discussed in subsection \ref{subsec:replica 2d} must be generalized to $\sigma_{\pm q}(\mu)$ in order to include a magnetic flux proportional to $\mu$. The conformal mapping which leads to the definition of the charged \ren entropy in a \emph{grand canonical ensemble} with an imaginary chemical potential is

\be\label{eq:unitary transform or rho grand canonical}
\tilde{\varrho}_\mt{A}^q(\mu_\mt{E}) \mapsto U^{-1}\tilde{\varrho}^q_\mt{A}(\beta_0,\mu_\mt{E})U =  U^{-1}\frac{e^{-q\(\beta_0 H-i\mu_\mt{E} Q_\mt{A}\)}}{\tilde{Z}^q(\beta_0,\mu_\mt{E})}U,
\ee
where $\tilde{Z}(\beta,\mu_\mt{E}) \equiv \Tr \left[e^{-\beta \(H-i\frac{\mu_\mt{E}}{\beta_0} Q_\mt{A}\)}\right]$ is the usual grand partition function. As before $U$ is a unitary transformation, which, upon taking the trace of the above expression, cancels with its inverse to give
\be\label{eq:Tr unitary transform or rho grand canonical}
\Tr \tilde{\varrho}^q_\mt{A}(\mu_\mt{E}) = \frac{\tilde{Z}(q\beta_0,\mu_\mt{E})}{\tilde{Z}^q(\beta_0,\mu_\mt{E})}.
\ee
Therefore, equation \eqref{eq:charged_Renyi_entropy_definition_imaginary} gives
\be\label{eq:charged_Renyi_entropy d-cft}
\tilde{S}_q(\mu_\mt{E}) = \frac{1}{q-1}\left( q \ln\tilde{Z}(\beta_0,\mu_\mt{E}) - \ln\tilde{Z}(q\beta_0,\mu_\mt{E})\right).
\ee

The thermal entropy $\tilde{S}(T,\mu_\mt{E})$ in the grand canonical ensemble is given in terms of the temperature derivative of the free energy
\be\label{eq:S(T,mu)}
\tilde{S}(T,\mu_\mt{E}) = -\(\frac{\partial \tilde{F}(T,\mu_\mt{E})}{\partial T}\)_{\mu_\mt{E}} = \(\frac{\partial}{\partial T}\left( T\ln\tilde{Z}(T^{-1},\mu_\mt{E}) \right)\)_{\mu_\mt{E}}.
\ee
One therefore arrives at the following relation between charged \ren entropy and the thermal entropy
\begin{equation}
\tilde{S}_q \left( {\mu_\mt{E}} \right) = \frac{q}{{q - 1}}\frac{1}{{T_0 }}\int_{T_0/q}^{T_0 } \! {dT \ \tilde{S} \left( {T,\mu_\mt{E} } \right)} .
\label{eq:Renyi_replica_trick}
\end{equation}

Technically speaking \cite{Baez:2011b}, \ren entropy is the $q$-derivative (or to be more precise the $q^{-1}$-derivative),\footnote{The $q^{-1}$-derivative is defined analogously but with $q^{-1}$ replacing $q$.} also known as the Jackson derivative, of the negative free energy
\be\label{eq:q-deriv def of S_q}
\tilde{S}_q \left( {\mu_\mt{E}} \right)=-D_{q^{-1}}\tilde{F}(T,\mu_\mt{E})= -\(\frac{\partial \tilde{F}(T,\mu_\mt{E})}{\partial T}\)_{q^{-1}}.
\ee
Interestingly enough, it is now obvious why taking the limit $q \to 1$ is a good idea, that is how $q$-derivatives reduce to ordinary ones. Straightforwardly, equation \eqref{eq:q-deriv def of S_q} reduces to \eqref{eq:S(T,mu)} in the limit $q \to 1$
\be\label{eq:S_q to S(T) limit}
S_\mt{EE} := \lim_{q \to 1} \tilde{S}_q \left( {\mu_\mt{E}} \right) = \tilde{S}(T,\mu_\mt{E}).
\ee
The above analysis is similar in the case of a real chemical potential.

%%%-----------------------------------------------------------------------------------------------------------------------------------------------------------------
\subsection{Correlators of Generalized Twist Operators}
\label{subsec:Correlators of Generalized Twist Operators}

In this subsection, we reproduce some of the results of \cite{Belin:2013uta} that will be of interest to us. For details the reader is referred to the above paper, as well as to \cite{Hung:2011nu}. As it was discussed above, \ren entropy can be realized by the insertion of a twist operator at the entangling surface. In the previous subsection, it was shown that inserting a Wilson loop in a correlator has a consequence; one must now generalize the twist operators $\sigma_{\pm q}$, appearing in the replica trick discussed in subsection \ref{subsec:replica 2d}, to include a magnetic flux proportional to $\mu$. The idea now is that by considering the leading singularity in the correlator $\langle T_{\mu\nu}\sigma_q\rangle$, one may define a generalized notion of the conformal dimension for  $\sigma_{\pm q}$.

If one inserts the stress tensor at a perpendicular distance $y$ from $\sigma_q$, with $y$ much smaller when compared to the other scales defining the geometry of $\mathscr{S}$, then
\be\label{eq:leading singularity of <Tσ>}
\begin{split}
\langle T_{ab}\sigma_q\rangle &= -\frac{h_q}{2\pi}\frac{\delta_{ab}}{y^d}, ~~~ \langle T_{a \alpha}\sigma_q\rangle = 0, \\
\langle T_{\alpha\beta}\sigma_q\rangle &= -\frac{h_q}{2\pi}\frac{(d-1)\delta_{\alpha\beta}-d n_\alpha n_\beta}{y^d},
\end{split}
\ee
where $\alpha,\beta\in\{0,1\}$ and $a,b\in\{2,...,d-1\}$, while $n_\alpha$ is the unit vector directed orthogonally from the twist operator to the insertion of the stress tensor. The singularity is fixed up to the conformal dimension $h_q$, which, as shown in \cite{Hung:2011nu}, it can be expressed in terms of the energy density
\be\label{eq:energy den}
\mathcal{E}(T,\mu):= \frac{E(T,\mu)}{R^{d-1}V_{\mathds{H}^{d-1}}},
\ee
on $S^1\times\mathds{H}^{d-1}$, where $V_{\mathds{H}^{d-1}}$ is the regulated volume of the hyperbolic plane $\mathds{H}^{d-1}$. From the first law of thermodynamics we have that
\be\label{eq:first law thermodynamics den}
d\mathcal{E}=Td\mathcal{S}+T_0 \mu d\mathcal{Q},
\ee
where $\mathcal{S}$ denotes the thermal entropy density and $\mathcal{Q}$ the thermal charge density respectively, therefore the conformal dimension is given by
\be\label{eq:h_q in terms of energy den}
h_q(\mu) = \frac{q}{d-1} \frac{R^{d-1}}{T_0}\int_{(T_0/q,\mu)}^{(T_0,0)} \! dT \ \partial_T\mathcal{E}(T,\mu) =\frac{q}{d-1}\frac{R^{d-1}}{T_0}\(\mathcal{E}(T_0,0)-\mathcal{E}(T_0/q,\mu)\).
\ee
The first term $\mathcal{E}(T_0,0)$ arises because, under conformal transformations, the stress tensor does not transform as a primary field (rather a quasi primary) and its transformation law includes an anomalous contribution. This contribution is the higher dimensional analog of the Schwarzian derivative appearing in equation \eqref{eq:transformation of T} \cite{Hung:2011nu}. Furthermore, note that the arguments of \cite{Hung:2011nu}, which give rise to equation \eqref{eq:h_q in terms of energy den}, apply also when one considers the calculation with a background gauge potential \cite{Belin:2013uta}.

As explained in \cite{Belin:2013uta}, in the context of charged \ren entropies, one should also consider correlators of $\sigma_q(\mu)$ with the current operator $J_\mu$ of the CFT, which is associate with the global charge \eqref{eq:global charge}. The conservation law of $J_\mu$ gives us the leading singularity in the correlator with the generalized twist operator $\sigma_q(\mu)$
\be\label{eq:leadin singularity of <Jσ>}
\langle J_{\alpha}\sigma_q(\mu)\rangle = -\frac{i k_q(\mu)}{2\pi}\frac{\epsilon_{\alpha\beta}n^\beta}{y^{d-1}}, ~~~ \langle J_{a}\sigma_q\rangle = 0,
\ee
where $\epsilon_{\alpha\beta}$ denotes the volume form in the two-dimensional space intersecting $\mathscr{S}$ and $k_q(\mu)$ is called the \emph{magnetic response} since it characterizes the response of the current to the magnetic flux. According to \cite{Belin:2013uta}, one can determine the value of $k_q(\mu)$ by using \eqref{eq: z to w transformation} and the conformal mapping \eqref{eq:conf mapping from D to H}. Starting from
\be\label{eq:J in terms og Q charge den}
\langle J_{\tau_\mt{E}} \rangle_{S^1\times \mathds{H}^{d-1}} = -i \mathcal{Q}(q,\mu),
\ee
in order to connect this thermal charge density to the correlator with the twist operator, we must conformally map the $S^1\times \mathds{H}^{d-1}$ background to a flat metric on $\mathds{R}^d$ using the transformation in \eqref{eq: z to w transformation}. The current $J_\mu$, under a conformal mapping, transforms as
\be\label{eq:J transformation}
\langle J_\mu \sigma_q(\mu)\rangle_{\mathds{R}^d} = \Omega^{d-2}\frac{\partial {x'}^{\nu}}{\partial x^\mu}\langle J'_{\nu}\rangle_{S^1\times \mathds{H}^{d-1}},
\ee
where the conformal factor $\Omega$ is given by \eqref{eq:conf mapping from D to H}. The mapping \eqref{eq: z to w transformation} generates a spherical twist operator. If one takes the limit with the current insertion approaching the spherical twist operator, one recovers the leading singularity in equation \eqref{eq:leadin singularity of <Jσ>}. Finally, using \eqref{eq: z to w transformation} and the conformal mapping \eqref{eq:conf mapping from D to H}, we find that
\be
k_q(\mu) = 2 \pi q R^{d-1} \mathcal{Q}(q;\mu).
\ee

As it was shown in \cite{Hung:2011nu} and subsequently in \cite{Belin:2013uta},\footnote{In particular, \cite{Perlmutter:2014} proved the statements of \cite{Hung:2011nu}, and laid the foundation for calculations that were later done in \cite{Belin:2013uta} and also proved \eqref{eq:d_q h_q [9]} to be true for any field theory, holographic or not.} the conformal dimension $h_q(\mu)$ and the magnetic flux response $k_q(\mu)$ have an interesting universal property. Their derivative with respect to $q$ when $q \to 1$ takes a very simple form. In particular, in \cite{Hung:2011nu} they found that
\be\label{eq:d_q h_q [9]}
\partial_q h_q|_{q=1} = \frac{2}{d-1}\pi^{1-\frac{d}{2}}\Gamma\(\frac{d}{2}\)\tilde{C}_T,
\ee
where $\tilde{C}_T$ is given in \eqref{eq:central charge of T} and it is the central charge defined by the two-point function of the stress tensor. However, in \cite{Belin:2013uta} and consequently to our case, this universal behaviour does not explicitly apply to the conformal dimension of the generalized twist operators $\sigma_q(\mu)$, but it is related to an expansion around $q=1$ and $\mu=0$
\be\label{eq:d_q h_q Belin}
h_q(\mu) = \sum_{m,n}\frac{1}{m!n!} h_{mn}(q-1)^m \mu^n, \ \  \mathrm{where} \ \  h_{mn} :=  (\partial_q)^m(\partial_\mu)^n h_q(\mu)|_{q=1,\mu=0}.
\ee
With this definition, note that $h_{00}=0$, while $h_{10}$ is precisely the term $\partial_q h_q|_{q=1}$ in equation \eqref{eq:d_q h_q [9]}. Similarly, for the magnetic response we have that
\be\label{eq:d_q k_q Belin}
k_q(\mu) = \sum_{m,n}\frac{1}{m!n!} k_{mn}(q-1)^m \mu^n, \ \  \mathrm{where} \ \  k_{mn} :=  (\partial_q)^m(\partial_\mu)^n k_q(\mu)|_{q=1,\mu=0}.
\ee
%%%-----------------------------------------------------------------------------------------------------------------------------------------------------------------
\section{Holographic Computation of \ren Entropy}
\label{sec:holographic}

The goal of this paper, is to use the AdS/CFT correspondence to calculate the \ren entropy for a spherical entangling surface in the boundary conformal field theory. We do this for CFTs, whose holographic dual is Gauss-Bonnet gravity coupled with an electromagnetic field. The additional coupling of Gauss-Bonnet gravity allows the study of a broader class of boundary conformal field theories. Standard black hole thermodynamics tools \cite{Wald:1993nt,Jacobson:1993vj,Iyer:1994ys} allow for the calculation of the horizon entropy for these solutions. According to the AdS/CFT dictionary, this black hole entropy is identical to the thermal entropy of the boundary CFT on the hyperbolic cylinder $\mathds{R} \times \mathds{H}^{d-1}$ (see subsection \ref{subsec:Charged Ren Entropy}). Thus, the AdS/CFT correspondence and application of formula \eqref{eq:Renyi_replica_trick} allow the calculation of entanglement \ren entropies in the boundary CFT, for spherical entangling surfaces, based solely on quantities of the bulk theory.

One should not forget that, since we study CFTs in the grand canonical ensemble, the boundary CFT contains a conserved current. Therefore, according to the AdS/CFT dictionary, the bulk theory will be invariant under a gauge symmetry, being the local extension of the global symmetry corresponding to the conserved current of the CFT and hence, it will contain the relevant gauge field. In the following, it is assumed that the latter is a $U(1)$ symmetry. The holographic representation of the grand canonical ensemble considered above will then be a topological black hole, which is charged under this gauge field. The advantage of this approach is the fact that analytic solutions of topological black holes with hyperbolic horizons are already known for the considered bulk theory \cite{Cvetic:2001bk,Ge:2008ni,Anninos:2008sj}.

Instead of considering the entropy as a function of the temperature as in equation \eqref{eq:Renyi_replica_trick}, it is more convenient to consider it as a function of horizon radius, normalized by the AdS curvature scale $x$,
\begin{equation}
x \equiv \frac{{r_\mt{H} }}{{\tilde L}}.
\label{eq:x_definition}
\end{equation}
In this case the \ren entropy can be expressed as
\begin{equation}
S_q \left( {\mu,\lambda } \right) = \frac{q}{{q - 1}}\frac{1}{{T_0 }}\int_{x_q }^{x_1 } \! {dx \ S\left( {x;\lambda } \right) {\partial_x T\left( {x;\mu,\lambda } \right)}},  \\
\label{eq:Renyi_entropy_formula}
\end{equation}
where $\lambda$ is the Gauss-Bonnet coupling, as we will see in the following. By parts integration of \eqref{eq:Renyi_entropy_formula} is expected to simplify calculations, since usually the formula for the entropy of the black hole is simpler than the formula for its temperature.

The quantity $x_q$ satisfies the equation
\begin{equation}
T\left( {x_q ,\mu } \right) = \frac{{T_0 }}{q}.
\label{eq:xn_definition}
\end{equation}
%%%-----------------------------------------------------------------------------------------------------------------------------------------------------------------
\subsection{Einstein-Gauss-Bonnet-Maxwell Gravity}
\label{subsec:EGBM}

In the following, we consider that the bulk theory is characterized by Gauss-Bonnet higher derivative corrections to the action, namely the bulk action is
\be\label{eq:einstein_gauss_bonnet_maxwell_action}
 I = \frac{1}{{2\ell_P ^{d - 1} }}\int \! {d^{d + 1} x} \ \sqrt { - g} \left( {\frac{{d\left( {d - 1} \right)}}{{L^2 }} + R}
 { + \frac{{\lambda L^2 }}{{\left( {d - 2} \right)\left( {d - 3} \right)}}\mathcal{X}- \frac{{\ell_* ^2 }}{4}F_{\mu \nu } F^{\mu \nu } } \right),
\ee
where $\mathcal{X} \equiv R^2  - 4R_{\mu \nu } R^{\mu \nu }  + R_{\mu \nu \kappa \lambda } R^{\mu \nu \kappa \lambda }$  is the Gauss-Bonnet term.

This class of theories have an interesting feature. The central charges of the corresponding dual CFT take specific values, because of the curvature squared coupling. Specifically, in four dimensions it is true that \cite{Myers:2010jv}
\begin{align}
c &= {\pi ^2}{\left( {\frac{{\tilde L}}{{{\ell_P}}}} \right)^3}\left( {1 - 2\lambda {f_\infty }} \right) ,\\
a &= {\pi ^2}{\left( {\frac{{\tilde L}}{{{\ell_P}}}} \right)^3}\left( {1 - 6\lambda {f_\infty }} \right) .
\end{align}
$\tilde L$ denotes the curvature length scale of AdS space, while $f_\infty$ is defined as $f_\infty = L^2 / {\tilde L}^2$.

Since we desire to calculate the entanglement \ren entropies for the corrsponding CFTs, we would like to relate the Gauss-Bonnet coupling with the central charges in arbitrary dimensions. Two central charges are required for this purpose. Following \cite{Hung:2011nu}, we use the central charge that is related with
the leading singularity of the two-point function of the stress tensor \cite{Buchel:2009sk},
\begin{equation}\label{eq:central charge of T}
{{\tilde C}_T} = \frac{{{\pi ^{\frac{d}{2}}}}}{{\Gamma \left( {\frac{d}{2}} \right)}}{\left( {\frac{{\tilde L}}{{{\ell_P}}}} \right)^{d - 1}}\left( {1 - 2\lambda {f_\infty }} \right)
\end{equation}
and the central charge defined in \cite{Myers:2010tj,Myers:2010xs}, which obeys an interesting holographic c-theorem
\begin{equation}
{a_d}^* = \frac{{{\pi ^{\frac{d}{2}}}}}{{\Gamma \left( {\frac{d}{2}} \right)}}{\left( {\frac{{\tilde L}}{{{\ell_P}}}} \right)^{d - 1}}\left( {1 - 2\frac{{d - 1}}{{d - 3}}\lambda {f_\infty }} \right) .
\end{equation}

In this paper, the Gauss-Bonnet bulk theory is studied as a holographic dual of a family of CFTs. One should be careful to demand that the corresponding boundary theory does not contain negative energy excitations \cite{Hofman:2008ar}. This demand imposes an acceptable range for the Gauss-Bonnet coupling \cite{Camanho:2009,Buchel:2009sk},
\begin{equation}
 - \frac{{\left( {3d + 2} \right)\left( {d - 2} \right)}}{{4{{\left( {d + 2} \right)}^2}}} \le \lambda  \le \frac{{\left( {d - 2} \right)\left( {d - 3} \right)\left( {{d^2} - d + 6} \right)}}{{4{{\left( {{d^2} - 3d + 6} \right)}^2}}} .
\label{eq:lambda_causality_bound}
\end{equation}
The above range, in terms of the central charges of the boundary CFT, can be written as
\begin{equation}
\frac{{d\left( {d - 3} \right)}}{{{d^2} - 2d - 2}} \le \frac{{{{\tilde C}_T}}}{{{a_d}^*}} \le \frac{d}{2}.
\label{eq:central_charge_bound}
\end{equation}

Considering CFTs with holographic duals, the relevant bulk solutions are charged topological black holes with hyperbolic horizons. These solutions are the holographic dual of the grand canonical ensemble of the boundary CFT. Spherical charged black hole solutions with hyperbolic horizons are provided in the literature \cite{Cvetic:2001bk,Ge:2008ni,Anninos:2008sj} (see also \cite{Cai:2002} for the uncharged solution). The solutions are of the form
\begin{equation}
d{s^2} =  - \left( {\frac{{{r^2}}}{{{L^2}}}f\left( r \right) - 1} \right)d{t^2} + \frac{{d{r^2}}}{{\frac{{{r^2}}}{{{L^2}}}f\left( r \right) - 1}} + {r^2}d{H _{d - 1}^2} ,
\label{eq:black_hole_metric}
\end{equation}
where by $d{H _{d - 1}^2}$, we denote the unit metric on the $(d-1)$-dimensional hyperbolic space $\mathds{H}^{d-1}$. The function $f\left( r \right)$ is given by
\begin{equation}
f\left( r \right) = \frac{1}{{2\lambda }}\left( {1 - \sqrt {1 - 4\lambda \left[ {1 + \frac{{{L^2}}}{{d - 1}}\left( {\frac{m}{{{r^{d - 1}}}} - \frac{{{q^2}}}{{2\left( {d - 2} \right){r^{2d - 2}}}}} \right)} \right]} } \right) ,
\label{eq:black_hole_full_solution}
\end{equation}
where $m$ and $q$ are constants of integration related with the mass and the electric charge respectively of the black hole.

The asymptotic behavior of the solution is
\begin{equation}
{f_\infty } \equiv \mathop {\lim }\limits_{r \to \infty } f\left( r \right) = \frac{{1 - \sqrt {1 - 4\lambda } }}{{2\lambda }}.
\label{eq:f_infinity}
\end{equation}
Thus, the curvature scale $\tilde L$ of the AdS, as shown in equation \eqref{eq:black_hole_metric}, is $
{{\tilde L}^2} = \frac{{{L^2}}}{{{f_\infty }}} $, with $f_\infty$ given by \eqref{eq:f_infinity}.

The time coordinate can be rescaled like $ t \to \frac{{\tilde L}}{\sqrt{f_\infty}R} t$, so that the boundary metric is conformally equivalent to
\begin{equation}
d{s_\infty }^2 =  - d{t^2} + {R^2}dH_{d - 1}^2 .
\end{equation}

The solution for the gauge field is\footnote{This solution is valid for $d>2$. If one wants to consider the $d=2$ case, which is of great interest, since it corresponds holographically to $2d$-CFTs, a logarithmic gauge field solution can be found. Thus, the $d=2$ case requires a separate study. However, since the Gauss-Bonnet term is a topological surface term for $d\leq 3$, the study is going to provide the same results with the Einstein-Maxwell case. For this reason, the reader is encouraged to see \cite{Belin:2013uta} for the $d=2$ case.}
\begin{equation}
A = \left( {\frac{1}{{d - 2}}\frac{{\tilde Lq}}{{R{\ell _*}{r^{d - 2}}}} - \frac{\mu }{{2\pi R}}} \right) dt .
\end{equation}
The chemical potential is selected in order for the total potential to vanish at the horizon in order to avoid a conical singularity. Thus,
\begin{equation}\label{eq:mu in terms of q}
\mu  = \frac{{2\pi \tilde Lq}}{{\left( {d - 2} \right){\ell _*}{r^{d - 2}}}} .
\end{equation}

The mass parameter $m$ is connected with the horizon radius as
\begin{equation}\label{eq:m}
m(x;\mu,\lambda) = \left( {d - 1} \right){{\tilde L}^{d - 2}}\left( {\frac{{{x^d}}}{{{f_\infty }}} - \left( {1 - \frac{{d - 2}}{{2\left( {d - 1} \right)}}{{\left( {\frac{{\mu {\ell _*}}}{{2\pi \tilde L}}} \right)}^2}} \right){x^{d - 2}} + \lambda {f_\infty }{x^{d - 4}}} \right) .
\end{equation}
We remind the reader that we defined $x$ to be the ratio of the horizon radius to the curvature length scale of AdS.

Calculating the Gibbs free energy for the black hole solution as $I = E - {E_b} - TS - \mu Q$, one can recover the energy and charge of the black hole with purely thermodynamic arguments
\begin{align}
E &= {\left( {\frac{{\partial I}}{{\partial {\beta_\mt{H}}}}} \right)_\mu } - \frac{\mu }{{{\beta_\mt{H}}}}{\left( {\frac{{\partial I}}{{\partial \mu }}} \right)_{{\beta_\mt{H}}}} = \frac{{m{V_{\mathds{H}^{d-1}} }}}{{2{\ell _P}^{d - 1}}} ,\label{eq:black_hole_energy} \\
Q &=  - \frac{1}{{{\beta_\mt{H}}}}{\left( {\frac{{\partial I}}{{\partial \mu }}} \right)_{{\beta_\mt{H}}}} = \frac{{q{\ell _*}}}{{{\ell _P}^{d - 1}}}{V_{\mathds{H}^{d-1}} }, \label{eq:black_hole_charge}
\end{align}
where $V_{\mathds{H}^{d-1}}$ is the regulated volume of the hyperbolic plane $\mathds{H}^{d-1}$. This volume is of course divergent; equations \eqref{eq:black_hole_energy} and \eqref{eq:black_hole_charge} for example, require appropriate regularization in order to make sense. This kind of divergences are further discussed in \cite{Casini:2011kv}. The energy $E_b$ is the energy of the extremal (zero-temperature) uncharged black hole solution. The selection of this background solution or even further the option to regularize of the Gibbs free energy with the use of counterterms may have interest when other kind of phase transitions, such as Hawking-Page phase transitions, are considered. In our case, it is not of great interest, as we are interested in differences of energies, as for example in the calculation of the conformal weights of twist operators.

In order to holographically calculate the charged \ren entropies, we should specify the Hawking temperature and thermal entropy of these black hole, as shown in \eqref{eq:Renyi_entropy_formula}.
The Hawking temperature of these black holes is given by \cite{Cvetic:2001bk,Ge:2008ni,Anninos:2008sj}
\begin{equation}
T\left( {x;\mu,\lambda} \right) = \frac{T_0}{2{f_\infty }}\frac{1}{x}\frac{{d{x^4} - \left( {d - 2} \right)\left( {1 + \frac{{d - 2}}{{2\left( {d - 1} \right)}}{{\left( {\frac{{\mu {\ell_*}}}{{2\pi \tilde{L}}}} \right)}^2}} \right){f_\infty }{x^2} + \left( {d - 4} \right)\lambda {f_\infty }^2}}{{{x^2} - 2\lambda {f_\infty }}} .
\label{eq:black_hole_temperature}
\end{equation}

One should be careful when calculating the thermal entropy of a black hole in a theory with higher derivative terms, since it is not proportional to the event horizon area. The desired thermal entropy has to be calculated using Wald's formula \cite{Wald:1993nt,Jacobson:1993vj,Iyer:1994ys} or equivalently using thermodynamic arguments, as for equations \eqref{eq:black_hole_energy} and \eqref{eq:black_hole_charge}, and is found equal to
\begin{equation}
{S}\left( {x;\lambda} \right) = {V_{\mathds{H}^{d-1}} }{\left( {\frac{{\tilde L}}{{{\ell_P}}}} \right)^{d - 1}}2\pi \left( {x^{d - 1} - 2\frac{{d - 1}}{{d - 3}}\lambda {f_\infty }x^{d - 3}} \right) .
\label{eq:black_hole_thermal_entropy}
\end{equation}

The last element missing to calculate the desired \ren entropy using formula \eqref{eq:Renyi_entropy_formula} is the quantity $x_q$. Equation \eqref{eq:xn_definition}, which defines $x_q$, combined with equation \eqref{eq:black_hole_temperature}, results in the following quartic equation for $x_q$.
\begin{equation}
d\frac{{x_q^4}}{{{f_\infty }}} - 2\frac{{x_q^3}}{q} - \left( {d - 2} \right)\left( {1 + \frac{{d - 2}}{{2\left( {d - 1} \right)}}{{\left( {\frac{{\mu {\ell_*}}}{{2\pi \tilde{L}}}} \right)}^2}} \right)x_q^2 + 4\lambda {f_\infty }\frac{{x_q}}{q} + \left( {d - 4} \right)\lambda {f_\infty } = 0 .
\label{eq:xn_equation}
\end{equation}
The appropriate $x_q$ is the largest real solution of equation \eqref{eq:xn_equation}. The limits for vanishing chemical potential studied in \cite{Hung:2011nu} and vanishing Gauss-Bonnet coupling studied in \cite{Belin:2013uta} can be easily recovered.\footnote{Notice that $ T\left( {1;0,\lambda} \right) = T_0$, which justifies why $x_1 = 1$, when there is no chemical potential, as in \cite{Hung:2011nu} for example.}

%%%-----------------------------------------------------------------------------------------------------------------------------------------------------------------
\subsection{Holographic Charged \ren Entropies}
\label{subsec:Renyi}

The \ren entropy for the charged Einstein-Gauss-Bonnet case can be calculated on the basis of equation \eqref{eq:Renyi_entropy_formula}, using equations \eqref{eq:black_hole_temperature} and \eqref{eq:black_hole_thermal_entropy} after an integration by parts. The acquired result is
\begin{multline}
{S_q}\left( {\mu ,\lambda} \right) = {V_{\mathds{H}^{d-1}} }{\left( {\frac{{\tilde L}}{{{\ell_P}}}} \right)^{d - 1}}\frac{{ q}}{{q - 1}}\frac{\pi}{{d - 3}}\left\{ {\frac{{d - 3}}{{{f_\infty }}}\left( {x_1^d - x_q^d} \right)} \phantom{\({\left( {\frac{{\mu {\ell_*}}}{{2\pi \tilde{L}}}} \right)}^2\)} \right.\\
 - \left( {3 + \frac{{d - 2}}{2}{{\left( {\frac{{\mu {\ell_*}}}{{2\pi \tilde{L}}}} \right)}^2}} \right)\left( {x_1^{d - 2} - x_q^{d - 2}} \right) - \left( {d - 1} \right)\lambda {f_\infty }\left( {x_1^{d - 4} - x_q^{d - 4}} \right)\\
\left. {\phantom{\({\left( {\frac{{\mu {\ell_*}}}{{2\pi \tilde{L}}}} \right)}^2\)} + \left[ {d\left( {1 - 4\lambda } \right) + \frac{{{{\left( {d - 2} \right)}^2}}}{{\left( {d - 1} \right)}}{{\left( {\frac{{\mu {\ell_*}}}{{2\pi \tilde{L}}}} \right)}^2}} \right]\left( {\frac{{x_1^d}}{{x_1^2 - 2\lambda {f_\infty }}} - \frac{{x_q^d}}{{x_q^2 - 2\lambda {f_\infty }}}} \right)} \right\} .
\label{eq:holographic_Renyi_entropy_result}
\end{multline}
The latter reduces to the results of \cite{Hung:2011nu} for vanishing chemical potential and to the results of \cite{Belin:2013uta} for vanishing Gauss-Bonnet coupling.

Let us make a break here to find motivation for rewriting equation \eqref{eq:holographic_Renyi_entropy_result} in a slightly different form. Equation \eqref{eq:holographic_Renyi_entropy_result} (as well as the vanishing Gauss-Bonnet coupling limit presented in \cite{Belin:2013uta}) may give the false impression that this result cannot be consistent with the Ryu-Takayanagi formula for the holographic entanglement entropy. Ryu and Takayanagi have made a very interesting conjecture \cite{Ryu:2006bv,Ryu:2006ef} for the entanglement entropy of CFTs with holographic duals whose gravitational sector is described by Einstein gravity. The conjecture suggests that the entanglement entropy of a given region $A$ equals
\begin{equation}
{S_1}\left( A \right) = \frac{{{\rm{Area}}\left( {{m_\mt{A}}} \right)}}{{4G_N^{d + 2}}},
\label{eq:RTformula}
\end{equation}
where $m_\mt{A}$ is the minimal surface that is homologous to $A$, or in other words the union of $A$ and $m_\mt{A}$ are the boundary of a higher dimensional manifold. The constant $G_N^{d + 2}$ is the gravitational constant in $d+2$ dimensions. Quantum corrections to the formula above should alter the value of the gravitational constant. $a'$ corrections, which correspond to higher derivative terms in the classical bulk action, should result in the substitution of the area law, with a different function, which matches the Wald formula \cite{Wald:1993nt,Jacobson:1993vj,Iyer:1994ys} for the black hole entropy.

The Ryu-Takayanagi formula has a simple and beautiful motivation. The entanglement entropy ${S_1}\left( A \right)$ is defined by tracing out the region $A^c$, thus it is the entropy that an observer in $A$ (who does not have access to $A^c$) measures. In the perspective of an asymptotically AdS black hole solution, the event horizon of the black hole naturally separates space in two regions and observers outside the horizon do not have access to any information from the inside. Then the minimal surface is simply identical to the horizon, resulting in formula \eqref{eq:RTformula} being similar to the Bekenstein-Hawking formula.

Although a general proof of formula \eqref{eq:RTformula} is very difficult \cite{Fursaev:2006,Headrick:2010}, the holographic calculations for spherical entangling surfaces agree with it \cite{Hung:2011nu}. It is a fair question whether this agreement insists in the case of charged entropies. If formula \eqref{eq:holographic_Renyi_entropy_result} was consistent with the Ryu-Takayanagi formula, it should reproduce the thermal entropy of the black hole in the limit $q \to 1$. However, this seems impossible due to the direct dependence on the chemical potential.

On the other hand, it was shown at the end of subsection \ref{subsec:Charged Ren Entropy} that formula \eqref{eq:Renyi_replica_trick}, which was the basis of our calculation, is consistent with Ryu-Takayanagi formula, as a simple consequence of $q$-derivatives reducing to ordinary ones in the limit $q \to 1$. We recall the result here
\be\label{eq:S_q to S(T) limit recall}
S_\mt{EE} := \lim_{q \to 1} S_q \left( {\mu} \right) = S(T,\mu).
\ee

The fact that formula \eqref{eq:holographic_Renyi_entropy_result} looks inconsistent with the expected result is an artifact because of the by parts integration that preceded the final result. Using the equation specifying $x_q$ \eqref{eq:xn_equation} to eliminate the term containing the chemical potential, one can find another simpler expression for the \ren entropy that has directly the correct $q \to 1$ limit.
\begin{multline}
{S_q}\left( {\mu ,\lambda } \right) = V_{\mathds{H}^{d-1}}{\left( {\frac{{\tilde L}}{{{\ell_P}}}} \right)^{d - 1}}2\pi \frac{q}{q - 1}\frac{{d - 1}}{{d - 2}} \left[ {\frac{{x_1^d - x_q^d}}{{{f_\infty }}}} \right. - \lambda {f_\infty }\left( {x_1^{d - 4} - x_q^{d - 4}} \right)\\
\left. { - \frac{1}{{d - 1}}\left( {x_1^{d - 1} - \frac{{x_q^{d - 1}}}{q}} \right) - \frac{{2\lambda {f_\infty }}}{{d - 3}}\left( {x_1^{d - 3} - \frac{{x_q^{d - 3}}}{q}} \right)} \right].
\label{eq:holographic_Renyi_entropy_result_simplified}
\end{multline}
This still depends implicitly on $\mu$ through $x_1$ and $x_q$. The latter, for vanishing Gauss-Bonnet coupling, simplifies to
\begin{equation}
{S_q}\left( {\mu ,0} \right) = {V_{\mathds{H}^{d-1}} }{\left( {\frac{\tilde{L}}{{{\ell_P}}}} \right)^{d - 1}}2\pi\frac{{q}}{{q - 1}}\frac{{d - 1}}{{d - 2}} \left[ {x_1^d - x_q^d - \frac{1}{{d - 1}}\left( {x_1^{d - 1} - \frac{{x_q^{d - 1}}}{q}} \right)} \right] .
\end{equation}

Let us now check some interesting limits of the acquired \ren entropy, as they are described in subsection \ref{subsec:renyi_basics}. The entanglement entropy can be calculated taking the appropriate limit of formula \eqref{eq:holographic_Renyi_entropy_result_simplified} or simply applying equation \eqref{eq:S_q to S(T) limit recall}. It is found equal to
\begin{equation}
{S_1}\left( \mu ,\lambda \right) = {V_{\mathds{H}^{d-1}} }{\left( {\frac{{\tilde L}}{{{\ell_P}}}} \right)^{d - 1}}2\pi \left( {x_1^{d - 1} - 2\frac{{d - 1}}{{d - 3}}\lambda {f_\infty }x_1^{d - 3}} \right) .
\label{eq:holographic _entaglement_entropy_result}
\end{equation}

The limit $q \to 0$, which specifies the number of non-vanishing eigenvalues of the density matrix, goes as
\begin{equation}
{S_0} \sim V_{\mathds{H}^{d - 1}}\pi{\left( {\frac{2}{d}} \right)^d}{\left( {\frac{{\tilde L}}{{{\ell _P}}}}\frac{f_\infty}{q} \right)^{d - 1}},
\end{equation}
while the limit $q \to \infty$, which specifies the largest of the eigenvalues of the density matrix, turns  out to be equal to
\begin{multline}
{S_\infty}\left( {\mu ,\lambda} \right) = {V_{\mathds{H}^{d-1}} }{\left( {\frac{{\tilde L}}{{{\ell_P}}}} \right)^{d - 1}}2\pi\frac{{d - 1}}{{d - 2}}\\
 \times \left[ {\frac{{x_1^d - x_\infty^d}}{{{f_\infty }}} - \lambda {f_\infty }\left( {x_1^{d - 4} - x_\infty^{d - 4}} \right) - \frac{{x_1^{d - 1}}}{{d - 1}} - \frac{{2\lambda {f_\infty }x_1^{d - 3}}}{{d - 3}}} \right] ,
\end{multline}
where
\begin{multline}
x_\infty^2 = \frac{{{f_\infty }}}{{2d}}\left\{ {\left( {d - 2} \right)\left( {1 + \frac{{d - 2}}{{2\left( {d - 1} \right)}}{{\left( {\frac{{\mu {\ell_*}}}{{2\pi \tilde{L}}}} \right)}^2}} \right)} \phantom{\left\{ { + \sqrt {{{\left( {d - 2} \right)}^2}{{\left( {1 + \frac{{d - 2}}{{2\left( {d - 1} \right)}}{{\left( {\frac{{\mu {\ell_*}}}{{2\pi \tilde{L}}}} \right)}^2}} \right)}^2} - 4d\left( {d - 4} \right)\lambda } } \right\}}\right. \\
\left. { + \sqrt {{{\left( {d - 2} \right)}^2}{{\left( {1 + \frac{{d - 2}}{{2\left( {d - 1} \right)}}{{\left( {\frac{{\mu {\ell_*}}}{{2\pi \tilde{L}}}} \right)}^2}} \right)}^2} - 4d\left( {d - 4} \right)\lambda } } \right\} .
\label{eq:xinfty}
\end{multline}

The expressions for the \ren entropy are quite complicated, since they require as an intermediate result the solution of equation \eqref{eq:xn_equation}. So it is useful to study \ren entropy numerically. Since the black hole thermodynamics for $d=4$ have different behaviour than for larger dimensions, a fact that it is reflected in the zeroth order term of equation \eqref{eq:xn_equation} that is proportional to $d-4$, \ren entropy is plotted for $d=4$ and when necessary for $d=5$. In order to remove the dependence of \ren entropy from the regulator volume $V_{\mathds{H}^{d-1}}$, instead of plotting $S_q$, we proceed to plot the ratio of $S_q$ with entanglement entropy $\frac{S_q}{S_1}$.

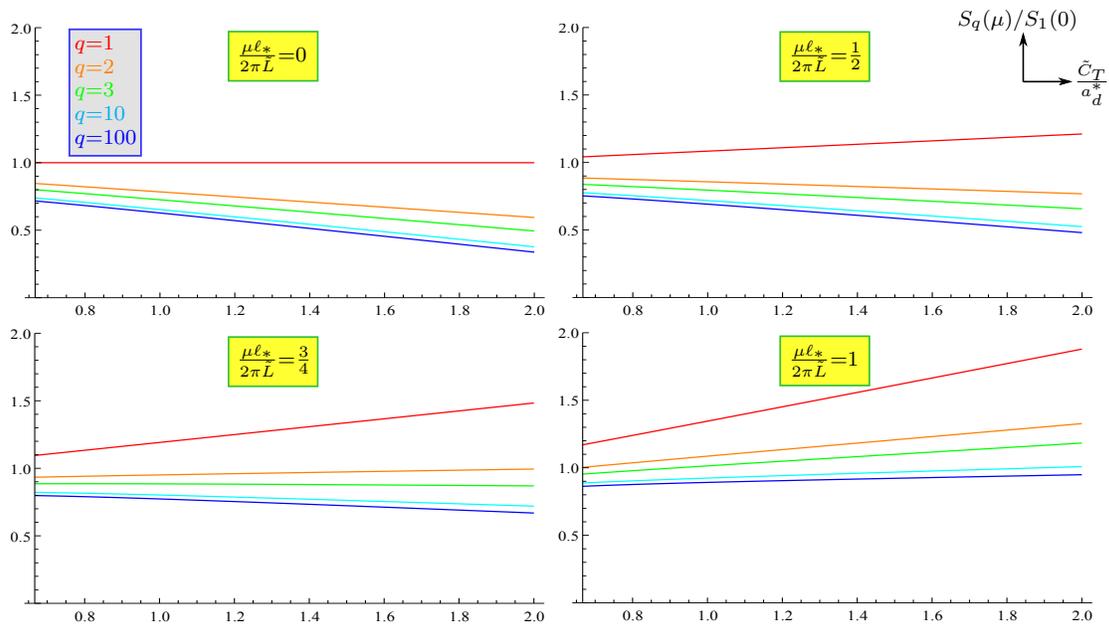
\begin{figure}[ht!]
\[
\raisebox{-225pt}{
  \begin{picture}(0,170)
  \put(-215,0){\scalebox{1.5}{\includegraphics{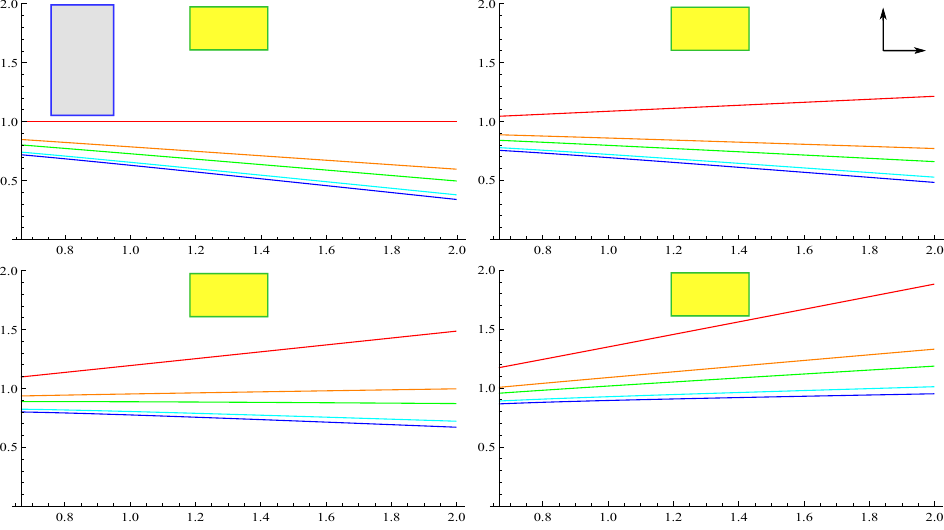}}}
  \put(0,5){
    \setlength{\unitlength}{.90pt}\put(-32,-1){
    \put(191,250) {$\scr S_q(\mu)/S_1(0)$}
    \put(240,224) {$\scr \frac{\tilde{C}_T}{a_d^*}$}
    \put(-180,240) {$\scr \color{red}q=1$}
    \put(-180,230) {$\scr \color{orange}q=2$}
    \put(-180,220) {$\scr \color{green}q=3$}
    \put(-180,210) {$\scr \color{cyan}q=10$}
    \put(-180,200) {$\scr \color{blue}q=100$}
    \put(-112,235) {$\scr \frac{\mu\ell_*}{2\pi \tilde{L}}=0$}
    \put(120,235) {$\scr \frac{\mu\ell_*}{2\pi \tilde{L}}=\frac{1}{2}$}
    \put(-112,107) {$\scr \frac{\mu\ell_*}{2\pi \tilde{L}}=\frac{3}{4}$}
    \put(120,107) {$\scr \frac{\mu\ell_*}{2\pi \tilde{L}}=1$}
      }\setlength{\unitlength}{1pt}}
  \end{picture}}
\]
\caption{\small{\ren entropies normalized by $S_1 \left( 0 \right)$ as function of the Gauss-Bonnet coupling for various chemical potentials for $d=4$}.}\label{Fig:3}
\end{figure}

\begin{figure}[ht!]
\[
\raisebox{-225pt}{
  \begin{picture}(0,170)
  \put(-215,0){\scalebox{1.5}{\includegraphics{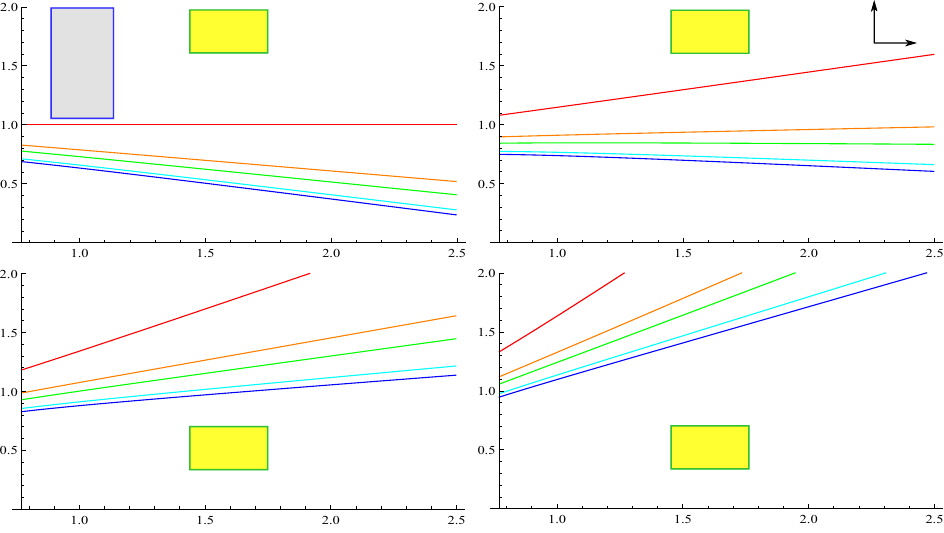}}}
  \put(0,5){
    \setlength{\unitlength}{.90pt}\put(-32,-1){
    \put(191,255) {$\scr S_q(\mu)/S_1(0)$}
    \put(240,230) {$\scr \frac{\tilde{C}_T}{a_d^*}$}
    \put(-180,240) {$\scr \color{red}q=1$}
    \put(-180,230) {$\scr \color{orange}q=2$}
    \put(-180,220) {$\scr \color{green}q=3$}
    \put(-180,210) {$\scr \color{cyan}q=10$}
    \put(-180,200) {$\scr \color{blue}q=100$}
    \put(-112,235) {$\scr \frac{\mu\ell_*}{2\pi \tilde{L}}=0$}
    \put(120,235) {$\scr \frac{\mu\ell_*}{2\pi \tilde{L}}=\frac{1}{2}$}
    \put(-112,34) {$\scr \frac{\mu\ell_*}{2\pi \tilde{L}}=\frac{3}{4}$}
    \put(120,34) {$\scr \frac{\mu\ell_*}{2\pi \tilde{L}}=1$}
      }\setlength{\unitlength}{1pt}}
  \end{picture}}
\]
\caption{\small{\ren entropies normalized by $S_1 \left( 0 \right)$ as function of the Gauss-Bonnet coupling for various chemical potentials for $d=5$}.}\label{Fig:4}
\end{figure}
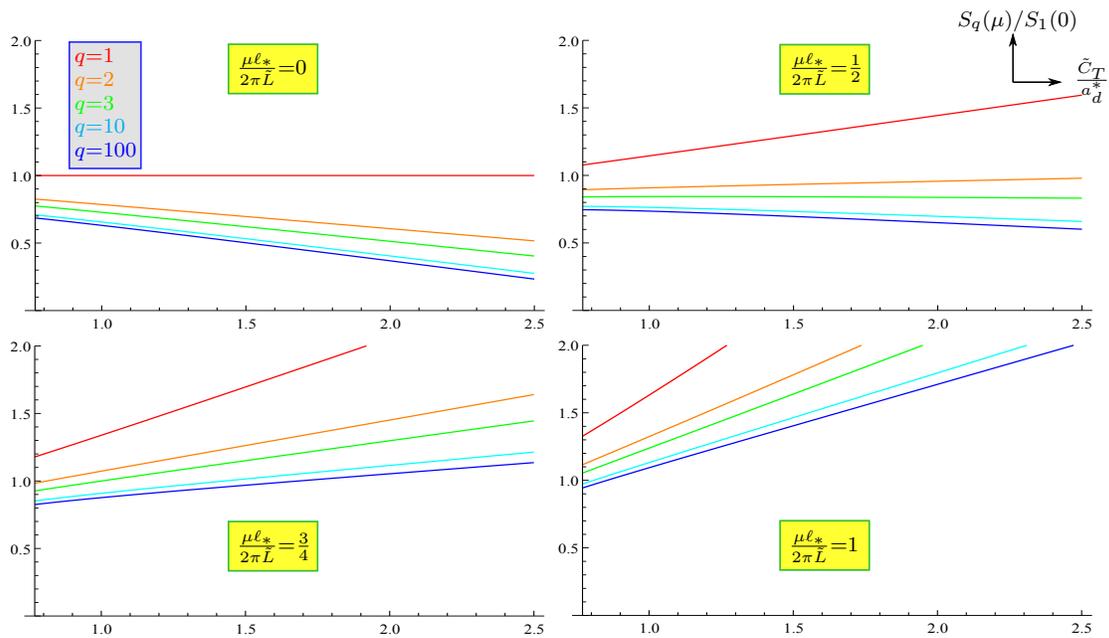

First, we plot the ratio $\frac{S_q \left( \mu \right)}{S_1 \left( 0 \right)}$ versus the central charges ratio $\frac{\tilde{C}_T}{a_d^*}$ for several different chemical potentials. The graphs are shown in figures \ref{Fig:3} and \ref{Fig:4}.

The plots are covering all the range of ratio $\frac{\tilde{C}_T}{a_d^*}$ allowed by causality as described by inequality \eqref{eq:central_charge_bound}. There are several observations to be made on these plots.

The plots are almost linear in the allowed by causality range of $\frac{\tilde{C}_T}{a_d^*}$. This is in agreement with the observations of \cite{Hung:2011nu}. However, as the chemical potential gets larger, the behaviour of the curves becomes increasingly non-linear.

Unlike the study for vanishing chemical potential \cite{Hung:2011nu}, the maximum value of the ratio $\frac{S_q \left( \mu \right)}{S_1 \left( 0 \right)}$ is not always achieved for the lower bound of the central charges ratio $\frac{\tilde{C}_T}{a_d^*} = \frac{d \left( d - 3 \right)}{d^2 - 2 d - 2}$. On the contrary, as the chemical potential increases, the ratio of $\frac{S_q \left( \mu \right)}{S_1 \left( 0 \right)}$ acquires its maximum value for the maximum bound of the central charges ratio $\frac{\tilde{C}_T}{a_d^*} = \frac{d}{2}$. For every $q$, the above statement is true only for chemical potentials larger than a critical value $\mu_{\mathrm{cr}} \left( q \right)$, that is an increasing function of $q$. However, above a critical chemical potential value, it becomes true for all $q$.

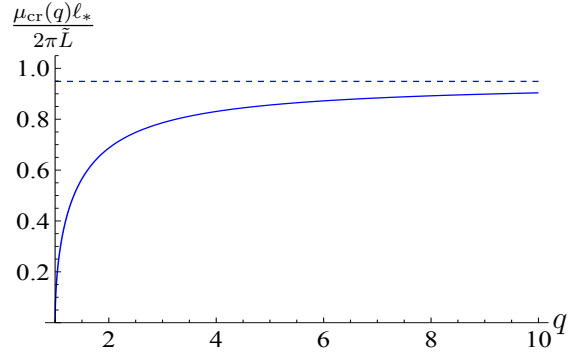
\begin{figure}[ht!]
\[
\raisebox{-58pt}{
  \begin{picture}(0,130)
  \put(-205,0){\scalebox{1.5}{\includegraphics{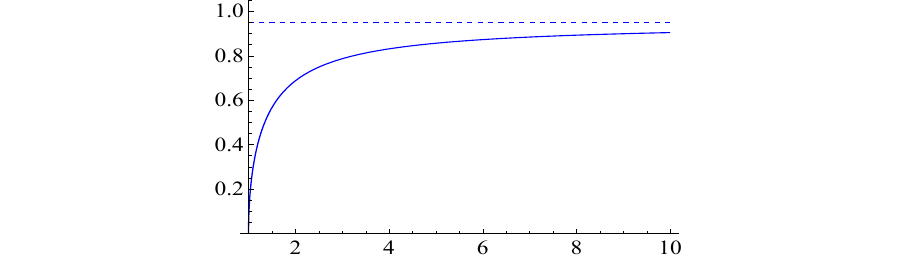}}}
  \put(0,5){
    \setlength{\unitlength}{.90pt}\put(-32,-1){
    \put(-95,128) {$\frac{\mu_{\mathrm{cr}}(q)\ell_*}{2 \pi \tilde{L}}$}
    \put(133,4) {$q$}
          }\setlength{\unitlength}{1pt}}
  \end{picture}}
\]
\caption{\small{The critical chemical potential for $d=4$, as a function of $q$.}}\label{fig:mucra}
\end{figure}

The above can be clarified more easily if one performs an expansion of \ren entropy around $\frac{\tilde{C}_T}{a_d^*} = 1$ and $\mu = 0$ that corresponds to the Einstein gravity case and it is exactly solvable. Such an expansion is performed in appendix \ref{sec:expansion}. Since the behavior of the \ren entropy curves are close to linear, the value of the coefficient of the linear term in $\frac{\tilde{C}_T}{a_d^*}$ is determining the behaviour of the curve and the position of the maximum. The calculation of the above coefficient up to second order in the chemical potential, suggests that the critical chemical potential as a function of $q$ for $d = 4$ can be estimated as
\begin{equation}
{\left( {\frac{{{\mu _{\mathrm{cr}}}\left( {q} \right){\ell _*}}}{{2\pi \tilde{L}}}} \right)^2} = \frac{{9\left( {(1 + 8{q^4} + \left( {1 - 4{q^2}} \right){{\left( {1 + 8{q^2}} \right)}^{\frac{1}{2}}}} \right)}}{{16{q^2}\left( { - 2 + 5{q^2} - \frac{{2 + 7{q^2}}}{{{{\left( {1 + 8{q^2}} \right)}^{\frac{1}{2}}}}}} \right)}} .
\label{eq:mucrd4}
\end{equation}
This is indeed an increasing function of $q$, as one can see in figure \ref{fig:mucra}.
In the figure, it is also visible that the limit $q \to \infty$ the critical potential reaches a finite value, which equation \eqref{eq:mucrd4} suggests being equal to ${\left( {\frac{{{\mu _{\mathrm{cr}}}\left( {\infty} \right){\ell _*}}}{{2\pi \tilde{L}}}} \right)^2} = \frac{9}{{10}}$. Thus, indeed for chemical potential larger than this value, $\frac{S_q \left( \mu \right)}{S_1 \left( 0 \right)}$ is an increasing function of the ratio of the central charges $\frac{\tilde{C}_T}{a_d^*}$ for all $q$.

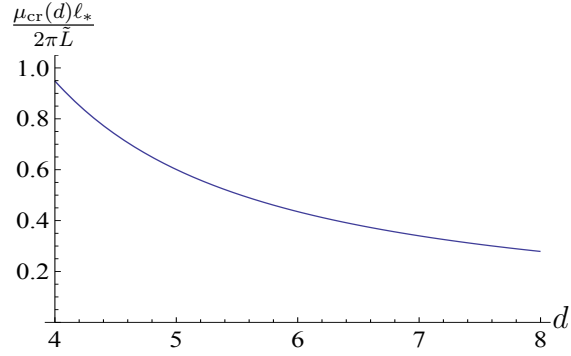
\begin{figure}[ht!]
\[
\raisebox{-58pt}{
  \begin{picture}(0,130)
  \put(-205,0){\scalebox{1.5}{\includegraphics{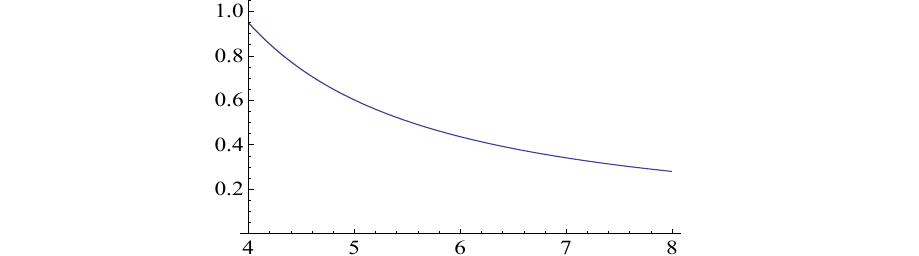}}}
  \put(0,5){
    \setlength{\unitlength}{.90pt}\put(-32,-1){
    \put(-95,128) {$\frac{\mu_{\mathrm{cr}}(d)\ell_*}{2 \pi \tilde{L}}$}
    \put(133,4) {$d$}
          }\setlength{\unitlength}{1pt}}
  \end{picture}}
\]
\caption{\small{The limit of the critical chemical potential when $q \to \infty$, as a function of the number of dimensions.}}\label{fig:mucrb}
\end{figure}

Similarly, one can find an expression for the critical chemical potential for any number of dimensions. If we restrict to the $q \to \infty$ limit, the complicated expressions are simplified quite a lot to find
\begin{equation}
{\left( {\frac{{{\mu _{\mathrm{cr}}}\left( {d} \right){\ell _*}}}{{2\pi \tilde{L}}}} \right)^2} = \frac{{4\left( {d - 1} \right)}}{{d{{\left( {d - 2} \right)}^2}\left( {{{\left( {\frac{{d - 2}}{d}} \right)}^{ - \frac{{d - 2}}{2}}} - 1} \right)}} ,
\end{equation}
which is a decreasing function of the number of dimensions, as it can be seen in figure \ref{fig:mucrb}.

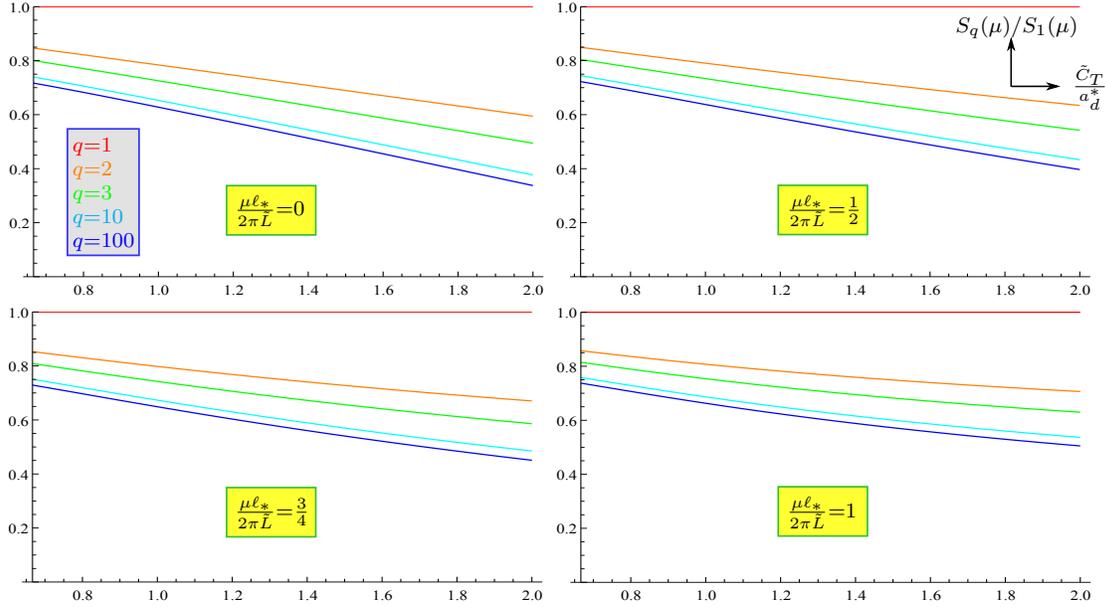
\begin{figure}[ht!]
\[
\raisebox{-225pt}{
  \begin{picture}(0,170)
  \put(-215,0){\scalebox{1.5}{\includegraphics{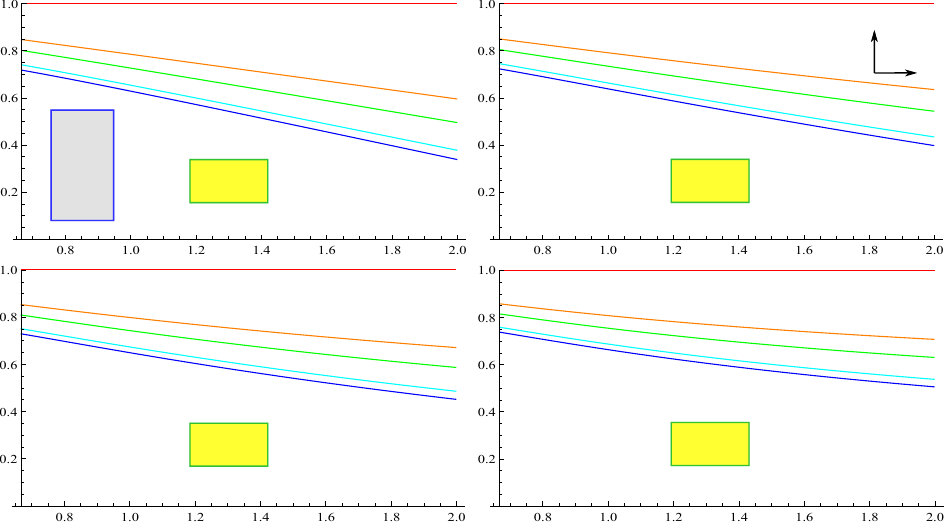}}}
  \put(0,5){
    \setlength{\unitlength}{.90pt}\put(-32,-1){
    \put(191,238) {$\scr S_q(\mu)/S_1(\mu)$}
    \put(240,213) {$\scr \frac{\tilde{C}_T}{a_d^*}$}
    \put(-180,188) {$\scr \color{red}q=1$}
    \put(-180,178) {$\scr \color{orange}q=2$}
    \put(-180,168) {$\scr \color{green}q=3$}
    \put(-180,158) {$\scr \color{cyan}q=10$}
    \put(-180,148) {$\scr \color{blue}q=100$}
    \put(-112,161) {$\scr \frac{\mu\ell_*}{2\pi \tilde{L}}=0$}
    \put(120,161) {$\scr \frac{\mu\ell_*}{2\pi \tilde{L}}=\frac{1}{2}$}
    \put(-112,34) {$\scr \frac{\mu\ell_*}{2\pi \tilde{L}}=\frac{3}{4}$}
    \put(120,34) {$\scr \frac{\mu\ell_*}{2\pi \tilde{L}}=1$}
      }\setlength{\unitlength}{1pt}}
  \end{picture}}
\]
\caption{\small{\ren entropies normalized by $S_1 \left( \mu \right)$ as function of the Gauss-Bonnet coupling for various chemical potentials for $d=4$}.}\label{Fig:5}
\end{figure}

Unlike the ratio $\frac{S_q \left( \mu \right)}{S_1 \left( 0 \right)}$, the ratio $\frac{S_q \left( \mu \right)}{S_1 \left( \mu \right)}$ always acquires its maximum value for the lower limit of the central charges ratio $\frac{\tilde{C}_T}{a_d^*} = \frac{d \left( d - 3 \right)}{d^2 - 2 d - 2}$, as shown in figure \ref{Fig:5}.

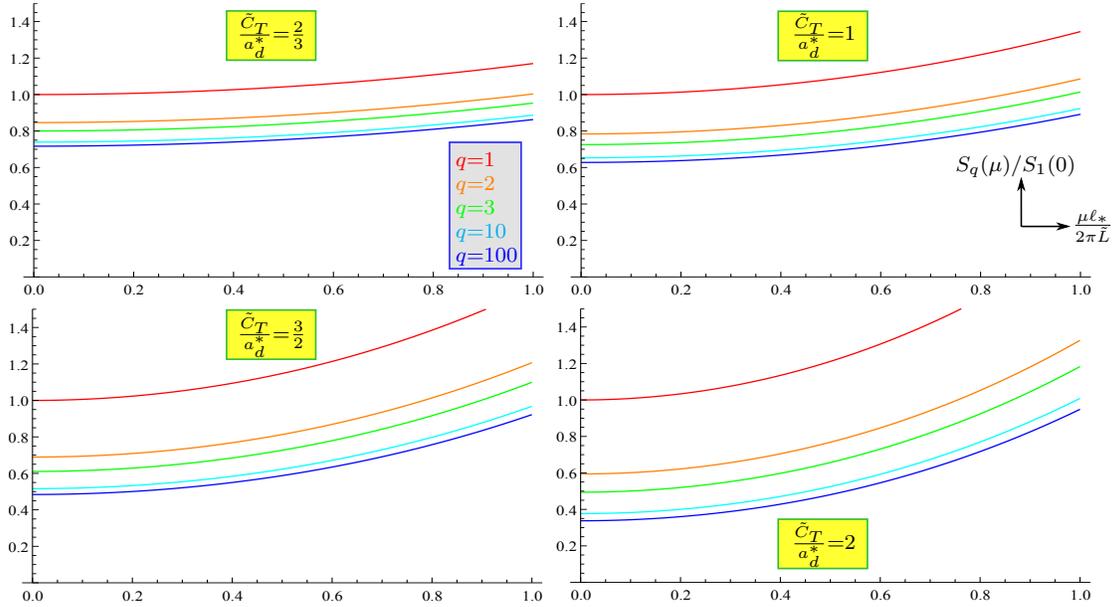
\begin{figure}[ht!]
\[
\raisebox{-225pt}{
  \begin{picture}(0,170)
  \put(-215,0){\scalebox{1.5}{\includegraphics{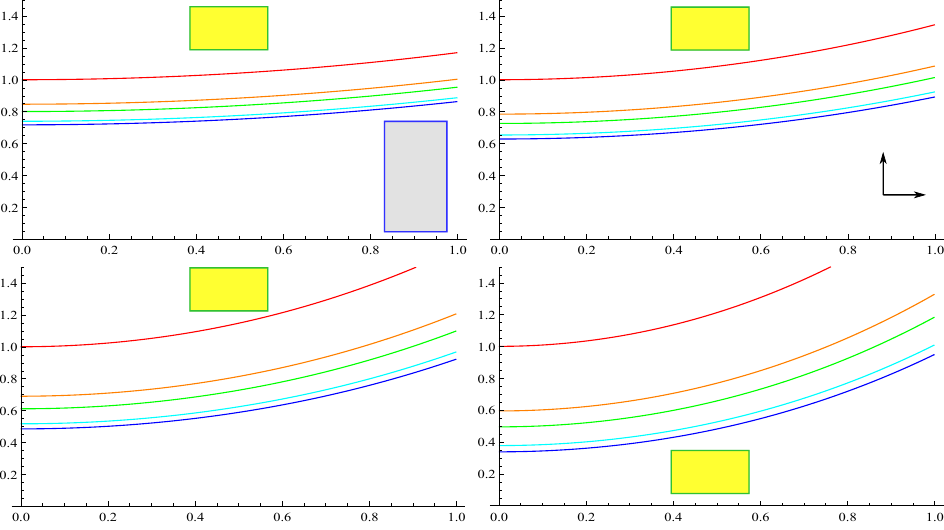}}}
  \put(0,5){
    \setlength{\unitlength}{.90pt}\put(-32,-1){
    \put(191,180) {$\scr S_q(\mu)/S_1(0)$}
    \put(240,155) {$\scr \frac{\mu\ell_*}{2\pi \tilde{L}}$}
    \put(-19,182) {$\scr \color{red}q=1$}
    \put(-19,172) {$\scr \color{orange}q=2$}
    \put(-19,162) {$\scr \color{green}q=3$}
    \put(-19,152) {$\scr \color{cyan}q=10$}
    \put(-19,142) {$\scr \color{blue}q=100$}
    \put(-111,235) {$\scr \frac{\tilde{C}_T}{a_d^*}=\frac{2}{3}$}
    \put(122,235) {$\scr \frac{\tilde{C}_T}{a_d^*}=1$}
    \put(-111,109) {$\scr \frac{\tilde{C}_T}{a_d^*}=\frac{3}{2}$}
    \put(122,21) {$\scr \frac{\tilde{C}_T}{a_d^*}=2$}
      }\setlength{\unitlength}{1pt}}
  \end{picture}}
\]
\caption{\small{\ren entropies normalized by $S_1 \left( 0 \right)$ as function of the chemical potential for various Gauss-Bonnet couplings for $d=4$.}}\label{Fig:7}
\end{figure}

We proceed to study the dependence of \ren entropy on the chemical potential. For this reason, in figure \ref{Fig:7} we plot the ratio $\frac{S_q \left( \mu \right)}{S_1 \left( 0 \right)}$ versus the chemical potential for several different values of the central charges ratio $\frac{\tilde{C}_T}{a_d^*}$.

The conclusion is that the specific ratio is always an increasing function of the chemical potential and actually this kind of dependence gets stronger for larger ratio of the central charges. This can also be understood in terms of the linear expansion of appendix \ref{sec:expansion}.

The dependence of $\frac{S_q \left( \mu \right)}{S_1 \left( 0 \right)}$ on the chemical potential is not linear, unlike its dependence on $\frac{\tilde{C}_T}{a_d^*}$. This is not unexpected, since only the square of the chemical potential appears in equation \eqref{eq:xn_equation} and thus, $x_q$ and \ren entropy depend only on even powers of the chemical potential. Indeed, if we plotted $\frac{S_q \left( \mu \right)}{S_1 \left( 0 \right)}$ versus the square of the chemical potential, we would see an almost linear dependence.

A very interesting feature of charged \ren entropies is the large chemical potential limit. Equation \eqref{eq:xn_equation} implies that for large chemical potentials, it is true that
\begin{equation}
x_q^2 \simeq {f_\infty }\frac{{{{\left( {d - 2} \right)}^2}}}{{2d\left( {d - 1} \right)}}{\left( {\frac{{\mu {\ell _*}}}{{2\pi \tilde{L}}}} \right)^2} .
\label{eq:xqlargemu}
\end{equation}
Interestingly enough, $x_q$ and therefore \ren entropy does not depend on $q$ at this limit. Specifically all \ren entropies tend to
\begin{equation}
\mathop {\lim }\limits_{\mu  \to \infty } {S_q}\left( {\mu ,\lambda} \right) \sim {V_{\mathds{H}^{d-1}}}2\pi{\left( {\frac{{\left( d - 2 \right) \sqrt{f_\infty}}}{{\sqrt {2d\left( {d - 1} \right)} }}\frac{{\mu {\ell _*}}}{{2\pi {\ell _P}}}} \right)^{d - 1}} ,
\end{equation}
It has to be noted that the limit depends only on the product $\sqrt{f_\infty} \mu$.

\begin{figure}[ht!]
\[
\raisebox{-225pt}{
  \begin{picture}(0,170)
  \put(-215,0){\scalebox{1.5}{\includegraphics{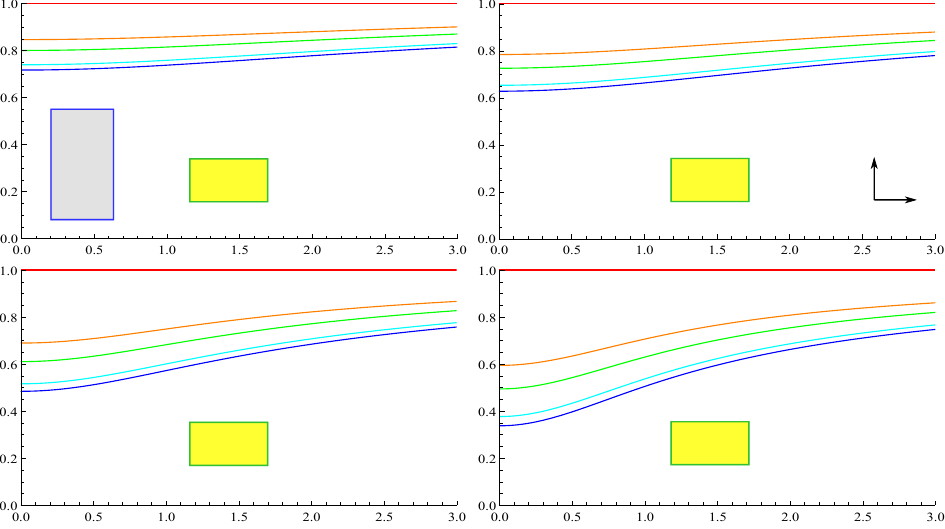}}}
  \put(0,5){
    \setlength{\unitlength}{.90pt}\put(-32,-1){
    \put(190,178) {$\scr S_q(\mu)/S_1(\mu)$}
    \put(237,152) {$\scr \frac{\mu\ell_*}{2\pi \tilde{L}}$}
    \put(-180,188) {$\scr \color{red}q=1$}
    \put(-180,178) {$\scr \color{orange}q=2$}
    \put(-180,168) {$\scr \color{green}q=3$}
    \put(-180,158) {$\scr \color{cyan}q=10$}
    \put(-180,148) {$\scr \color{blue}q=100$}
    \put(-113,162) {$\scr \frac{\tilde{C}_T}{a_d^*}=\frac{2}{3}$}
    \put(121,162) {$\scr \frac{\tilde{C}_T}{a_d^*}=1$}
    \put(-113,35) {$\scr \frac{\tilde{C}_T}{a_d^*}=\frac{3}{2}$}
    \put(123,35) {$\scr \frac{\tilde{C}_T}{a_d^*}=2$}
      }\setlength{\unitlength}{1pt}}
  \end{picture}}
\]
\caption{\small{\ren entropies normalized by $S_1 \left( \mu \right)$ as a function of the chemical potential for various Gauss-Bonnet couplings for $d=4$.}}\label{Fig:9}
\end{figure}
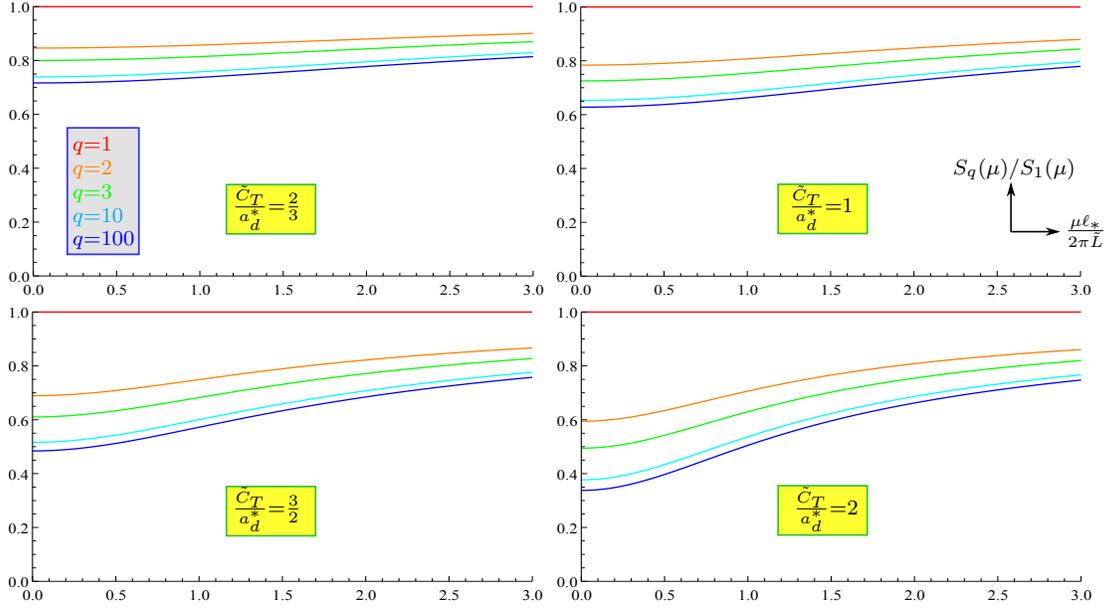

The independence of $S_q$ and $q$ for large chemical potentials is made clear in figure \ref{Fig:9} where $\frac{S_q \left( \mu \right)}{S_1 \left( \mu \right)}$ is plotted versus the chemical potential. Clearly all curves converge to one for large chemical potentials.

As mentioned in subsection \ref{subsec:Charged Ren Entropy}, the study of \ren entropy for imaginary chemical potential may reveal interesting features related to confinement. It is known \cite{Witten:1998zw} that AdS/CFT correspondence connects the confinement/deconfinement phase transition of the boundary conformal field theory with the Hawking-Page phase transition \cite{Hawking:1982dh} of the dual bulk theory. More specifically, the high temperature black hole phase in the bulk is dual to the deconfining phase in the boundary theory, while the low temperature thermal AdS phase in the bulk is dual to the confining phase in the boundary theory. However, it is also known \cite{Anninos:2008sj,Cvetic:2001bk} that asymptotically AdS black holes with hyperbolic horizons never undergo a Hawking-Page phase transition in any number of dimensions. Thus, it is not expected to discover any discontinuity in the periodicity of $\tilde{S}_q$.

A first thing to notice on the behaviour of \ren entropy for imaginary chemical potential is the qualitative difference between the case $\frac{\tilde{C}_T}{a_d^*}\geq1$ (non-negative Gauss-Bonnet coupling) and the case $\frac{\tilde{C}_T}{a_d^*}<1$ (negative Gauss-Bonnet coupling). In figures \ref{Fig:11} and \ref{Fig:12}, where the ratio $\frac{{{{\tilde S}_q}\left( {{\mu_\mt{E}} } \right)}}{{{{\tilde S}_1}\left( {{0} } \right)}}$ is  plotted versus the magnitude of the imaginary chemical potential for $d=4$ and $d=5$ respectively.
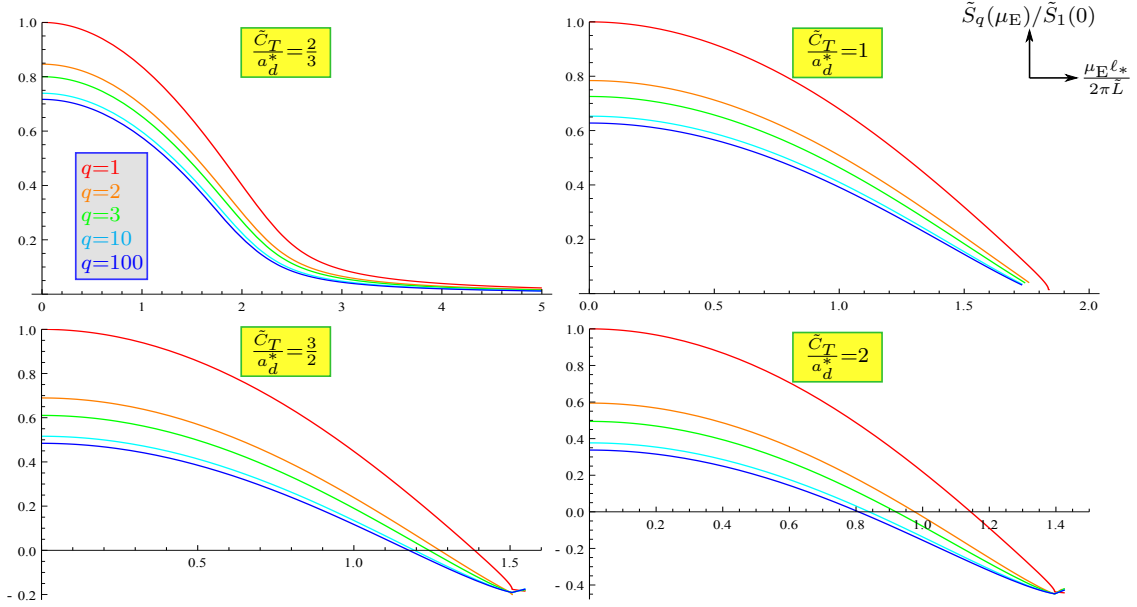
\begin{figure}[ht!]
\[
\raisebox{-225pt}{
  \begin{picture}(0,170)
  \put(-215,0){\scalebox{1.5}{\includegraphics{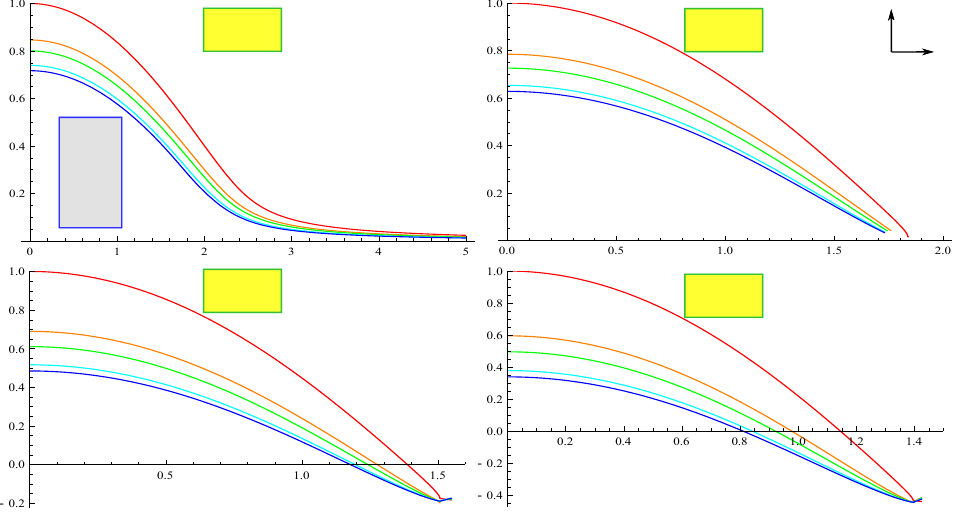}}}
  \put(0,5){
    \setlength{\unitlength}{.90pt}\put(-32,-1){
    \put(194,250) {$\scr \tilde{S}_q(\mu_\mt{E})/\tilde{S}_1(0)$}
    \put(244,224) {$\scr \frac{\mu_\mt{E}\ell_*}{2\pi \tilde{L}}$}
    \put(-176,186) {$\scr \color{red}q=1$}
    \put(-176,176) {$\scr \color{orange}q=2$}
    \put(-176,166) {$\scr \color{green}q=3$}
    \put(-176,156) {$\scr \color{cyan}q=10$}
    \put(-176,146) {$\scr \color{blue}q=100$}
    \put(-105,235) {$\scr \frac{\tilde{C}_T}{a_d^*}=\frac{2}{3}$}
    \put(128,235) {$\scr \frac{\tilde{C}_T}{a_d^*}=1$}
    \put(-105,109) {$\scr \frac{\tilde{C}_T}{a_d^*}=\frac{3}{2}$}
    \put(128,107) {$\scr \frac{\tilde{C}_T}{a_d^*}=2$}
      }\setlength{\unitlength}{1pt}}
  \end{picture}}
\]
\caption{\small{\ren entropies normalized by $\tilde{S}_1 \left( 0 \right)$ as a function of the imaginary chemical potential for various Gauss-Bonnet couplings for $d=4$.}}\label{Fig:11}
\end{figure}
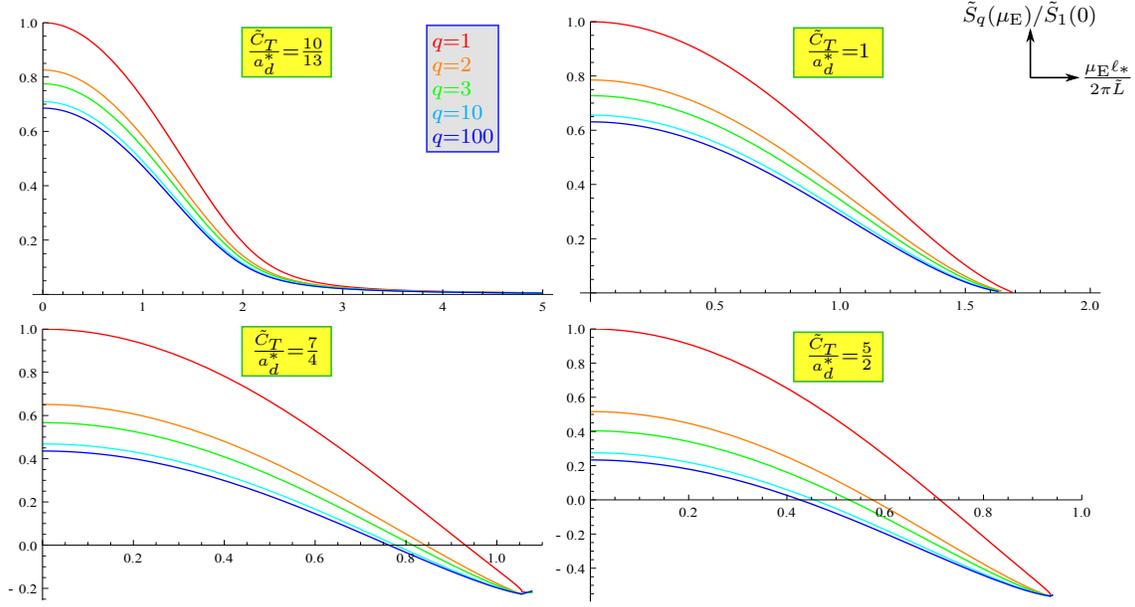
\begin{figure}[ht!]
\[
\raisebox{-225pt}{
  \begin{picture}(0,170)
  \put(-215,0){\scalebox{1.5}{\includegraphics{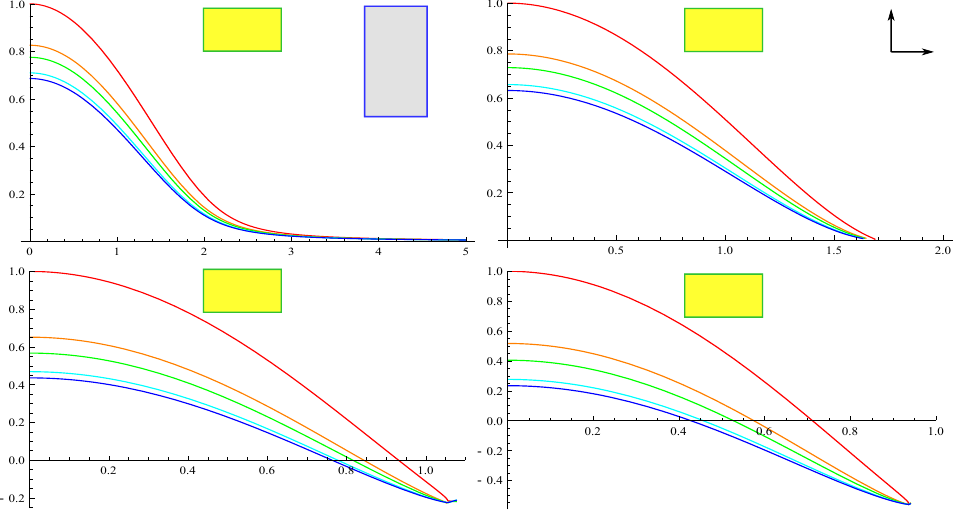}}}
  \put(0,5){
    \setlength{\unitlength}{.90pt}\put(-32,-1){
    \put(194,250) {$\scr \tilde{S}_q(\mu_\mt{E})/\tilde{S}_1(0)$}
    \put(244,224) {$\scr \frac{\mu_\mt{E}\ell_*}{2\pi \tilde{L}}$}
    \put(-29,239) {$\scr \color{red}q=1$}
    \put(-29,229) {$\scr \color{orange}q=2$}
    \put(-29,219) {$\scr \color{green}q=3$}
    \put(-29,209) {$\scr \color{cyan}q=10$}
    \put(-29,199) {$\scr \color{blue}q=100$}
    \put(-107,235) {$\scr \frac{\tilde{C}_T}{a_d^*}=\frac{10}{13}$}
    \put(128,235) {$\scr \frac{\tilde{C}_T}{a_d^*}=1$}
    \put(-105,109) {$\scr \frac{\tilde{C}_T}{a_d^*}=\frac{7}{4}$}
    \put(128,107) {$\scr \frac{\tilde{C}_T}{a_d^*}=\frac{5}{2}$}
      }\setlength{\unitlength}{1pt}}
  \end{picture}}
\]
\caption{\small{\ren entropies normalized by $\tilde{S}_1 \left( 0 \right)$ as a function of the imaginary chemical potential for various Gauss-Bonnet couplings for $d=5$.}}\label{Fig:12}
\end{figure}

For both $d=4$ and $d=5$, in the case $\frac{\tilde{C}_T}{a_d^*} < 1 $, \ren entropies are well defined for arbitrarily large imaginary chemical potential. Moreover, the ratio $\frac{{{{\tilde S}_q}\left( {{\mu_\mt{E}} } \right)}}{{{{\tilde S}_1}\left( {{0} } \right)}}$ goes to zero for $\mu_\mt{E}$ going to infinity. On the other hand in the case $\frac{\tilde{C}_T}{a_d^*} \geq 1 $, the curves are interrupted at a finite value of the imaginary chemical potential $\mu_{\mt{Ecr}} \left(\frac{\tilde{C}_T}{a_d^*} , q \right)$. Technically this happens, because equation \eqref{eq:xn_equation} fails to have a real and positive solution. Physically, as the imaginary chemical potential increases further than the bound $\mu_{\mt{Ecr}}$ the black hole with temperature equal to $T = \frac{T_0}{q}$ turns to a naked singularity.

It is quite complicated to calculate the exact form of $\mu_{\mt{Ecr}} \left(\frac{\tilde{C}_T}{a_d^*} , q \right)$, since this requires the study of the roots of a quartic equation. However, equation \eqref{eq:xn_equation} becomes quadratic for vanishing Gauss-Bonnet coupling, in which case it is very simple to find that
\begin{equation}
{\left( {\frac{{{\mu_{\mt{Ecr}}}\left( {1,q} \right){\ell _*}}}{{2\pi \tilde{L}}}} \right)^2} = \frac{{2\left( {d - 1} \right)}}{{d - 2}}\left( {1 - \frac{1}{{d\left( {d - 2} \right){q^2}}}} \right).
\end{equation}
Indeed, this equation agrees with the top right parts of figures \ref{Fig:11} and \ref{Fig:12}, which show that the larger the $q$, the smaller the $\mu_{\mt{Ecr}}\left( {1,q} \right)$. This trend gets inverted for larger values of $\frac{\tilde{C}_T}{a_d^*} \geq 1 $, thus enforcing all \ren entropy curves to get interrupted at the same imaginary chemical potential, since for the definition of any $S_q$, $x_1$ is required.

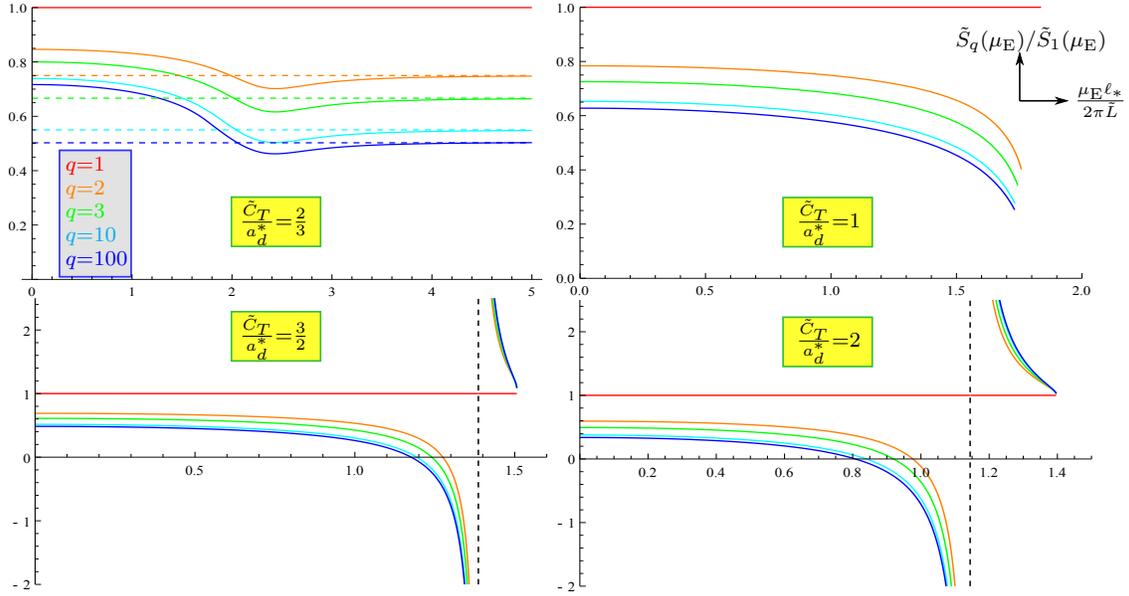
\begin{figure}[ht!]
\[
\raisebox{-225pt}{
  \begin{picture}(0,170)
  \put(-215,0){\scalebox{1.5}{\includegraphics{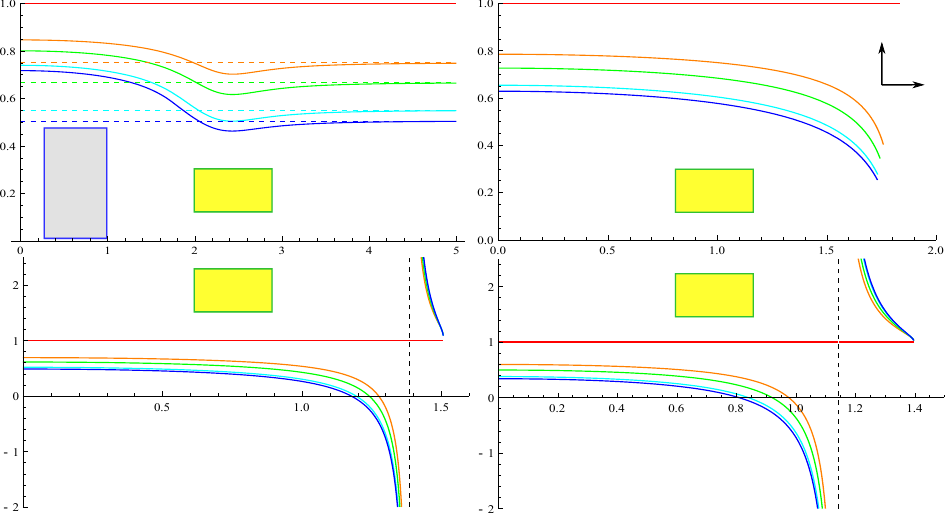}}}
  \put(0,5){
    \setlength{\unitlength}{.90pt}\put(-32,-1){
    \put(191,233) {$\scr \tilde{S}_q(\mu_\mt{E})/\tilde{S}_1(\mu_\mt{E})$}
    \put(241,208) {$\scr \frac{\mu_\mt{E}\ell_*}{2\pi \tilde{L}}$}
    \put(-183,181) {$\scr \color{red}q=1$}
    \put(-183,171) {$\scr \color{orange}q=2$}
    \put(-183,161) {$\scr \color{green}q=3$}
    \put(-183,151) {$\scr \color{cyan}q=10$}
    \put(-183,141) {$\scr \color{blue}q=100$}
    \put(-110,157) {$\scr \frac{\tilde{C}_T}{a_d^*}=\frac{2}{3}$}
    \put(124,157) {$\scr \frac{\tilde{C}_T}{a_d^*}=1$}
    \put(-110,109) {$\scr \frac{\tilde{C}_T}{a_d^*}=\frac{3}{2}$}
    \put(124,107) {$\scr \frac{\tilde{C}_T}{a_d^*}=2$}
      }\setlength{\unitlength}{1pt}}
  \end{picture}}
\]
\caption{\small{\ren entropies normalized by $\tilde{S}_1 \left( \mu_\mt{E} \right)$ as a function of the imaginary chemical potential for various Gauss-Bonnet couplings for $d=4$. Each dashed line is for $S_q=\frac{q+1}{2q}$ with each $q$ given by the respective colour.}}\label{Fig:13}
\end{figure}

\begin{figure}[ht!]
\[
\raisebox{-225pt}{
  \begin{picture}(0,170)
  \put(-215,0){\scalebox{1.5}{\includegraphics{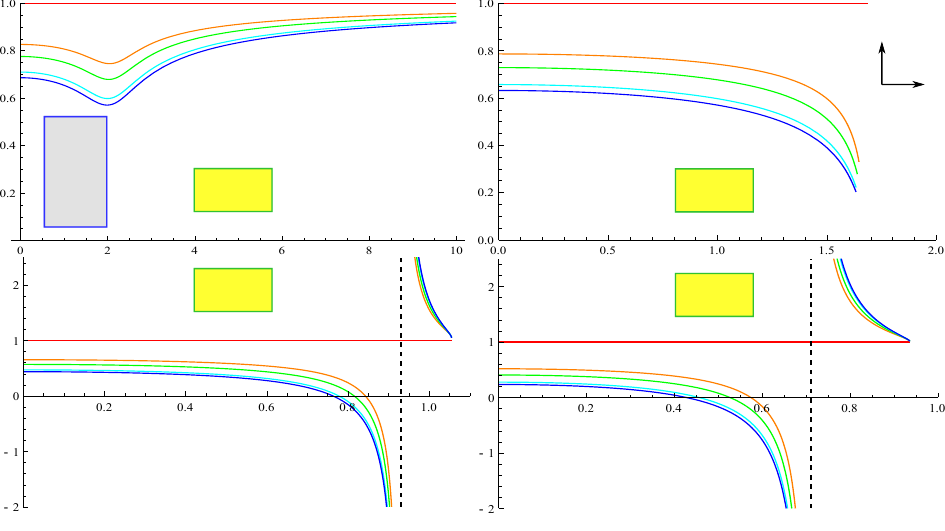}}}
  \put(0,5){
    \setlength{\unitlength}{.90pt}\put(-32,-1){
    \put(191,233) {$\scr \tilde{S}_q(\mu_\mt{E})/\tilde{S}_1(\mu_\mt{E})$}
    \put(241,208) {$\scr \frac{\mu_\mt{E}\ell_*}{2\pi \tilde{L}}$}
    \put(-183,186) {$\scr \color{red}q=1$}
    \put(-183,176) {$\scr \color{orange}q=2$}
    \put(-183,166) {$\scr \color{green}q=3$}
    \put(-183,156) {$\scr \color{cyan}q=10$}
    \put(-183,146) {$\scr \color{blue}q=100$}
    \put(-112,157) {$\scr \frac{\tilde{C}_T}{a_d^*}=\frac{10}{13}$}
    \put(124,157) {$\scr \frac{\tilde{C}_T}{a_d^*}=1$}
    \put(-110,109) {$\scr \frac{\tilde{C}_T}{a_d^*}=\frac{7}{4}$}
    \put(124,107) {$\scr \frac{\tilde{C}_T}{a_d^*}=\frac{5}{2}$}
      }\setlength{\unitlength}{1pt}}
  \end{picture}}
\]
\caption{\small{\ren entropies normalized by $\tilde{S}_1 \left( \mu_\mt{E} \right)$ as a function of the imaginary chemical potential for various Gauss-Bonnet couplings for $d=5$.}}\label{Fig:14}
\end{figure}

In figures \ref{Fig:13} and \ref{Fig:14}, we plot $\frac{{{{\tilde S}_q}\left( {{\mu_\mt{E}} } \right)}}{{{{\tilde S}_1}\left( {{\mu} } \right)}}$ versus the magnitude of the imaginary chemical potential for $d=4$ and $d=5$ respectively.

As long as the $\frac{\tilde{C}_T}{a_d^*} < 1 $ is concerned, an interesting observation is that the asymptotic value of $\frac{{{{\tilde S}_q}\left( {{\mu_\mt{E}} } \right)}}{{{{\tilde S}_1}\left( {{\mu} } \right)}}$ is different for $d=4$ than in $d=5$ case. This is due to the fact that especially for $d=4$, the zeroth order term of equation \eqref{eq:xn_equation} vanishes. This fact discriminates the asymptotic behaviour of the largest solution of \eqref{eq:xn_equation} for $d=4$ and $d>4$. More specifically
\begin{equation}
\lim_{\mu_\mt{E}\to\infty}x_q \sim \begin{cases}
  - \cfrac{{6\lambda {f_\infty }}}{q}{\left( {\cfrac{{{\mu_\mt{E}}{\ell _*}}}{{2\pi \tilde{L}}}} \right)^{ - 2}}, & d = 4 ,\\
\cfrac{{\sqrt { - 2\left( {d - 1} \right)\left( {d - 4} \right)\lambda {f_\infty }} }}{{\left( {d - 2} \right)}}{\left( {\cfrac{{{\mu_\mt{E}}{\ell _*}}}{{2\pi \tilde{L}}}} \right)^{ - 1}} , & d > 4 .
\end{cases}
\end{equation}
In both cases $x_q$ tends to zero for large imaginary potentials and more interestingly, for any $d > 4$ it does not depend on $q$. For $d = 4$ the dominant term of \ren entropy is the one proportional to $x_1-\frac{x_q}{q}$. The asymptotic behavior of \ren entropy is
\begin{equation}
{{\tilde S}_q}\left( {{\mu_\mt{E}},\lambda } \right) \sim
\begin{cases}
{V_{\mathds{H}^{d-1}}}{\left( {\cfrac{{\tilde L}}{{{\ell _P}}}} \right)^3}36\pi{\left( {\lambda {f_\infty }} \right)^2}{\left( {\cfrac{{{\mu_\mt{E}}{\ell _*}}}{{2\pi \tilde{L}}}} \right)^{ - 2}}\cfrac{q + 1}{q}, & d = 4 ,\\
 V_{\mathds{H}^{d-1}}{\left( {\cfrac{{\tilde L}}{{{\ell _P}}}} \right)^{d - 1}} \cfrac{\pi}{d - 3} {\left[ { - 2\left( {d - 1} \right)\lambda {f_\infty }} \right]^{\frac{{d - 1}}{2}}}{\left( {\cfrac{{d - 2 }}{{\sqrt {d - 4}}}} {\cfrac{{{\mu_\mt{E}}{\ell _*}}}{{2\pi \tilde L}}} \right)^{ - \left( {d - 3} \right)}}, & d > 4 ,
\end{cases}
\end{equation}
resulting in the discrimination between the $d = 4$ case and $d > 4$ that is clearly visible in the top left part of figures \ref{Fig:13} and \ref{Fig:14}, namely
\begin{equation}
\mathop {\lim }\limits_{{\mu_\mt{E}} \to \infty } \frac{{{{\tilde S}_q}\left( {{\mu_\mt{E}},\lambda } \right)}}{{{{\tilde S}_1}\left( {{\mu_\mt{E}},\lambda } \right)}} = \begin{cases} \cfrac{{q + 1}}{{2q}}, & d = 4 , \\
1, & d > 4 .
\end{cases}
\end{equation}

As long as the $\frac{\tilde{C}_T}{a_d^*} > 1 $ is concerned, first we note the existence of a vertical asymptote of the curves. This is simply due to the fact that ${{{{\tilde S}_1}\left( {{\mu_\mt{E}} } \right)}}$ vanishes at some finite $\mu_\mt{E} = {{{\left. {{\mu_\mt{E}}} \right|}_{{S_1} = 0}}}$. The value of $ {{{\left. {{\mu_\mt{E}}} \right|}_{{S_1} = 0}}}$ is not difficult to be specified. Equation \eqref{eq:holographic _entaglement_entropy_result} suggest that this happens when $x_1^2 = 2\frac{{d - 1}}{{d - 3}}\lambda {f_\infty }$. Then using equation \eqref{eq:xn_equation} it is not difficult to find
\begin{multline}
{\left( {\frac{{{{\left. {{\mu_\mt{E}}} \right|}_{{S_1} = 0}}{\ell _*}}}{{2\pi \tilde{L}}}} \right)^2} = \frac{{2\left( {d - 1} \right)}}{{d - 2}}\\
 - \frac{1}{{{{\left( {d - 2} \right)}^2}}}\left( {\frac{{4d{{\left( {d - 1} \right)}^2}\lambda }}{{d - 3}} - 8\sqrt {\frac{{2\left( {d - 1} \right)}}{{d - 3}}\lambda {f_\infty }}  + \left( {d - 3} \right)\left( {d - 4} \right)} \right) .
\end{multline}

%%%-----------------------------------------------------------------------------------------------------------------------------------------------------------------
\subsection{\ren Entropy Inequalities}\label{subsec:ren entropy inequalities}

As discussed in subsection \ref{subsec:renyi_basics}, \ren entropies obey a series of interesting inequalities. These inequalities are direct consequence of the fact that \ren entropies are defined on a probability distribution, in our case the spectrum of the density matrix $\rho_\mt{A}$. However, when this calculation is performed taking advantage of holographic dualities, the probability distribution is not accessible, as the calculation is based on black hole thermodynamics. However, if the holographic theory is meant to have a CFT dual at a stable thermal ensemble, it must still obey the \ren entropy inequalities. This can impose interesting restrictions in the bulk theory. Following the analysis in \cite{Hung:2011nu}, we proceed to identify the properties of the black holes in the bulk theory that are required in order for the \ren entropy inequalities to be valid.

Starting from equation \eqref{eq:Renyi_replica_trick}, one can show that
\begin{equation}
\frac{{\partial {S_q \left( \mu \right)}}}{{\partial q}} =  - \frac{1}{{{{\left( {q - 1} \right)}^2}}}\frac{1}{{{T_0}}}\int_{\frac{{{T_0}}}{q}}^{{T_0}} {\left[ {S\left( T, \mu \right) - S\left( {\frac{{{T_0}}}{q}} ,\mu \right)} \right]dT} .
\end{equation}
The specific heat is given by
\begin{equation}
{C_\mu } = T{\left( {\frac{{\partial S \left(T, \mu \right)}}{{\partial T}}} \right)_\mu } .
\end{equation}
Assuming that the black holes under consideration are thermally stable, in others words, they are characterized by positive specific heat, the integrated quantity in the above integral is positive when $q > 1$, and negative when $q < 1$. However, the upper limit of integration is larger than the lower limit when $q > 1$ and smaller when $q < 1$, resulting in the integral being positive for every $q$. Thus, the first inequality holds when the black holes under consideration are thermally stable.

Similarly, from equation \eqref{eq:Renyi_replica_trick}, one can show that
\begin{equation}
\frac{\partial }{{\partial q}}\left( {\frac{{q - 1}}{q}{S_q \left( \mu \right)}} \right) = \frac{1}{{{q^2}}}S\left( {\frac{{{T_0}}}{q}} , \mu \right) .
\label{eq:renyi_second_inequality_holographic}
\end{equation}
This means that the second inequality holds as long as the black holes under consideration are characterized by positive thermal entropy.

In the same way, one can find that
\begin{equation}
\frac{\partial }{{\partial q}}\left( {\left( {q - 1} \right){S_q \left( \mu \right) }} \right) = S\left( {\frac{{{T_0}}}{q}} , \mu \right) + \frac{1}{{{T_0}}}\int_{\frac{{{T_0}}}{q}}^{{T_0}} {\left[ {S\left( T , \mu \right) - S\left( {\frac{{{T_0}}}{q}} , \mu \right)} \right]dT} .
\end{equation}
The integral in the above equation is positive for every $q$, as long as thermal stability is ensured, as shown in the proof of the first inequality. Thus, the third inequality holds if the black holes under consideration are both thermally stable and are characterized by positive thermal entropy.

Finally, one can show that
\begin{equation}
\frac{{{\partial ^2}}}{{\partial {q^2}}}\left( {\left( {q - 1} \right){S_q \left( \mu \right)}} \right) =  - \frac{{{T_0}}}{{{q^3}}}{\left. {\frac{{\partial S \left( T , \mu \right)}}{{\partial T}}} \right|_{T = \frac{{{T_0}}}{q}}} =  - \frac{1}{{{q^2}}}{C_\mu }\left( {\frac{{{T_0}}}{q}} , \mu \right) ,
\end{equation}
meaning that the forth inequality holds, as long as the black holes under consideration are thermally stable.

It turns out that in the case of charged black holes, similarly to the uncharged ones, \ren entropy inequalities are satisfied, as long as the topological black holes, interfering in the holographic calculation of \ren entropy are thermodynamically stable and are characterized by positive thermal entropy. Let' s specify the restrictions to the bulk theory that are imposed by these requirements.

In \cite{Anninos:2008sj}, it was shown that in Gauss-Bonnet gravity, charged black holes with hyperbolic horizons may be characterized by negative thermal entropy or negative specific heat. Specifically, negative entropy black holes appear only for Gauss-Bonnet couplings larger than
\begin{equation}
\lambda  > \frac{{\left( {d - 3} \right)\left( {{d^2} + d - 8} \right)}}{{4d{{\left( {d - 1} \right)}^2}}}
\label{eq:lambda_negative_entropy_bound}
\end{equation}
and only for chemical potentials obeying
\begin{equation}
{{{\left( {\frac{{\mu {\ell _*}}}{{2\pi \tilde{L}}}} \right)}^2}} < \frac{1}{{{{\left( {d - 2} \right)}^2}}}\left( {\frac{{d{{\left( {d - 1} \right)}^2}}}{{d - 3}}\lambda  - \frac{{{d^2} + d - 8}}{4}} \right) .
\label{eq:chemical_negative_entropy}
\end{equation}
Negative specific heat black holes always have negative entropy and appear only when the Gauss-Bonnet coupling is larger than
\begin{equation}
\lambda  > \frac{1}{4}
\end{equation}
and only for chemical potentials obeying
\begin{equation}
{{{\left( {\frac{{\mu {\ell _*}}}{{2\pi \tilde{L}}}} \right)}^2}} < \frac{{d\left( {d - 1} \right)}}{{{{\left( {d - 2} \right)}^2}}}\left( {\lambda  - \frac{1}{4}} \right) .
\end{equation}

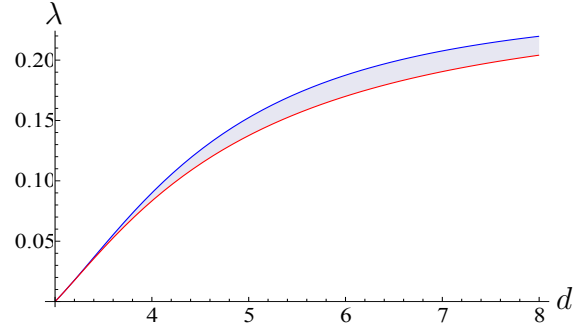
\begin{wrapfigure}{r}{0.5\textwidth}
\vspace{-15pt}
\[
\raisebox{-58pt}{
  \begin{picture}(0,140)
  \put(-110,0){\scalebox{1.5}{\includegraphics{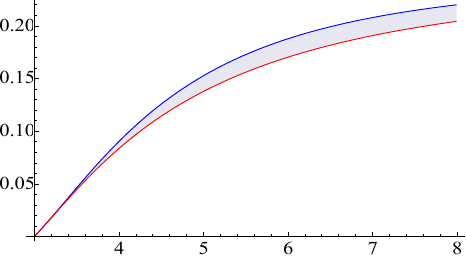}}}
  \put(0,5){
    \setlength{\unitlength}{.90pt}\put(-32,-1){
    \put(-78,123) {$\lambda$}
    \put(137,3) {$d$}
          }\setlength{\unitlength}{1pt}}
  \end{picture}}
\]
\vspace{-15pt}
\caption{\small{The bounds imposed on the Gauss-Bonnet coupling by causality and the positivity of the thermal entropy of black holes as function of the number of dimensions.}}\label{fig:lambdabounds}
%\vspace{-10pt}
\end{wrapfigure}

All figures of previous subsection are clearly in agreement with the first inequality. This is not unexpected since asymtotically AdS black holes with hyperbolic horizons are always thermally stable in any number of dimensions, as long as the Gauss-Bonnet coupling is smaller than $\frac{1}{4}$, which is required by causality as imposed by equation \eqref{eq:lambda_causality_bound}. However, this is not the case for black holes with negative thermal entropy. As shown in figure \ref{fig:lambdabounds}, for $d>3$, there is always a range for the Gauss-Bonnet coupling that does not violate the causality bounds, while it allows for black holes with negative thermal entropy.

\noindent
Couplings below the blue curve do not violate causality, couplings above the red curve allow for negative entropy black holes, while the shaded region corresponds to coupling obeying both properties.

\begin{figure}[ht!]
\[
\raisebox{-58pt}{
  \begin{picture}(0,130)
  \put(-215,0){\scalebox{1.5}{\includegraphics{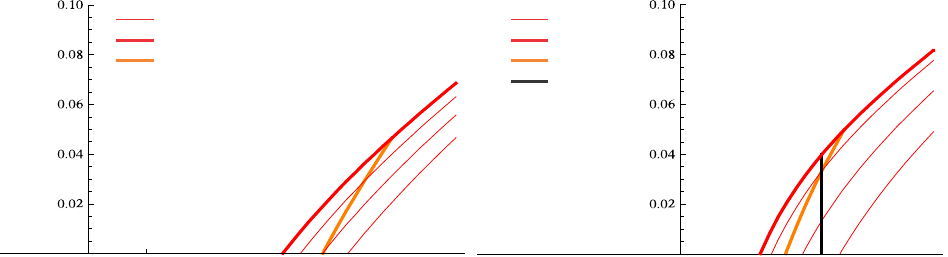}}}
  \put(0,5){
    \setlength{\unitlength}{.90pt}\put(-32,-1){
    \put(-170,125) {$T$}
    \put(-130,108) {$\scr\mu=\mathrm{const}$}
    \put(-130,98) {$\scr\mu=0$}
    \put(-130,88) {$\scr m=0$}
    \put(60,108) {$\scr\mu=\mathrm{const}$}
    \put(60,98) {$\scr\mu=0$}
    \put(60,88) {$\scr m=0$}
    \put(60,78) {$\scr S=0$}
    \put(115,125) {$T$}
    \put(250,-5) {$x$}
    \put(-160,-15) {\tiny{$\scr\sqrt{2 \frac{d-1}{d-3} \lambda f_\infty}$}}
    \put(165,-15) {\tiny{$\scr\sqrt{2 \frac{d-1}{d-3} \lambda f_\infty}$}}
          }\setlength{\unitlength}{1pt}}
  \end{picture}}
\]
\caption{\small{Black hole thermodynamics. On the l.h.s for $\lambda  < \frac{{\left( {d - 3} \right)\left( {{d^2} + d - 8} \right)}}{{4d{{\left( {d - 1} \right)}^2}}}$ and on the r.h.s for $\lambda  > \frac{{\left( {d - 3} \right)\left( {{d^2} + d - 8} \right)}}{{4d{{\left( {d - 1} \right)}^2}}}$ \cite{Anninos:2008sj}.}}\label{fig:bhthermodynamics}
\end{figure}

Equation \eqref{eq:black_hole_thermal_entropy} suggests that black holes characterized by negative thermal entropy must obey
\begin{equation}
{x^2} < x_{S = 0}^2 \equiv 2\frac{{d - 1}}{{d - 3}}\lambda {f_\infty } .
\end{equation}
Then, equation \eqref{eq:black_hole_temperature} implies that black holes with negative thermal entropy have Hawking temperature smaller than
\begin{equation}
T\left( {{x_{S = 0}},\mu } \right) = \frac{{{T_0}}}{8}\frac{1}{{\sqrt {2\frac{{d - 1}}{{d - 3}}\lambda {f_\infty }} }}\left[ {\frac{{4d{{\left( {d - 1} \right)}^2}}}{{d - 3}}\lambda  - \left( {{d^2} + d - 8} \right) - {{\left( {d - 2} \right)}^2}{{\left( {\frac{{\mu {\ell _*}}}{{2\pi \tilde{\tilde{L}}}}} \right)}^2}} \right] .
\end{equation}
Black hole thermodynamics are depicted in figure \ref{fig:bhthermodynamics}.

\begin{figure}[ht!]
\[
\raisebox{-58pt}{
  \begin{picture}(0,130)
  \put(-215,0){\scalebox{1.5}{\includegraphics{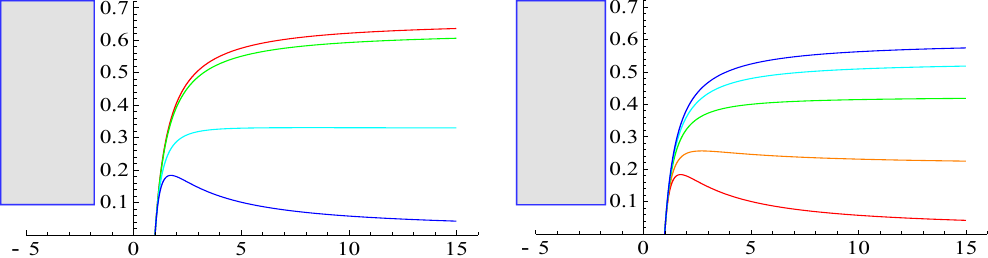}}}
  \put(0,5){
    \setlength{\unitlength}{.90pt}\put(-32,-1){
    \put(-170,130){$\frac{q-1}{q}\frac{S_q(0)}{S_1(0)}$}
    \put(77,130)  {$\frac{q-1}{q}\frac{S_q(\mu)}{S_1(\mu)}$}
    \put(269,4)   {$q$}
    \put(24,4)    {$q$}
    \put(-207,104){\color{red}$\frac{\tilde{C}_T}{a_d^*}=\frac{28}{33}$}
    \put(-207,79) {\color{green}$\frac{\tilde{C}_T}{a_d^*}=1$}
    \put(-207,54) {\color{cyan}$\frac{\tilde{C}_T}{a_d^*}=\frac{9}{4}$}
    \put(-207,29) {\color{blue}$\frac{\tilde{C}_T}{a_d^*}=\frac{7}{2}$}
    \put(47,108)  {\color{blue}$\scr \frac{\mu\ell_*}{2\pi \tilde{L}}=1$}
    \put(47,88)   {\color{cyan}$\scr \frac{\mu\ell_*}{2\pi \tilde{L}}=\frac{3}{4}$}
    \put(47,68)   {\color{green}$\scr \frac{\mu\ell_*}{2\pi \tilde{L}}=\frac{1}{2}$}
    \put(47,48)   {\color{orange}$\scr \frac{\mu\ell_*}{2\pi \tilde{L}}=\frac{1}{4}$}
    \put(47,28)   {\color{red}$\scr \frac{\mu\ell_*}{2\pi \tilde{L}}=0$}
          }\setlength{\unitlength}{1pt}}
  \end{picture}}
\]
\caption{\small{The violation of second inequality obeyed by \ren entropies for $d=7$.}}\label{fig:inequality_violation}
\end{figure}
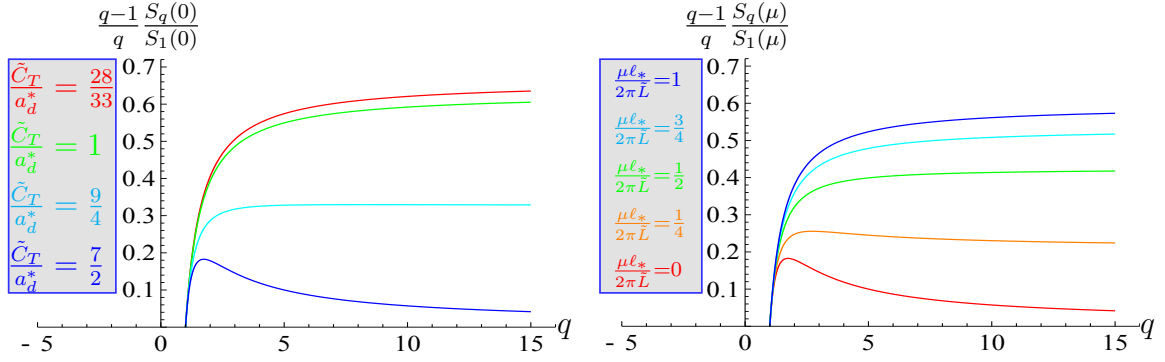

According to the above, for chemical potentials obeying \eqref{eq:chemical_negative_entropy}, it is expected, that for high enough $q$, $\frac{T_0}{q}$ will become small enough to correspond to a negative entropy black hole. In this case, for large $q$, it is expected that equation \eqref{eq:renyi_second_inequality_holographic} will give rise to a violation of the second inequality for \ren entropies. Specifically this is expected to occur for any $q > q_v$, where
\begin{equation}
{q_v} = \frac{{{T_0}}}{{T\left( {{x_{S = 0}},\mu } \right)}} = \frac{{8\sqrt {2\frac{{d - 1}}{{d - 3}}\lambda {f_\infty }} }}{{\frac{{4d{{\left( {d - 1} \right)}^2}}}{{d - 3}}\lambda  - \left( {{d^2} + d - 8} \right) - {{\left( {d - 2} \right)}^2}{{\left( {\frac{{\mu {\ell _*}}}{{2\pi \tilde{L}}}} \right)}^2}}}
\end{equation}
Indeed, this is the case as it can be seen in figure \ref{fig:inequality_violation}.
On the left, $\frac{q-1}{q}\frac{S_q}{S_1}$ is plotted versus $q$ for vanishing chemical potential and for several Gauss-Bonnet couplings covering all the allowed by causality region. On the right, $\frac{q-1}{q}\frac{S_q}{S_1}$ is plotted versus $q$ for the maximum value of Gauss-Bonnet coupling allowed by causality and for several different chemical potentials.

%%%-----------------------------------------------------------------------------------------------------------------------------------------------------------------
\subsection{Holographic Calculation of  \texorpdfstring{$h_q(\mu,\lambda)$}{h} and  \texorpdfstring{$k_q(\mu,\lambda)$}{k}}\label{subsec:Holographic Calculation of h and k}

In subsection \ref{subsec:Correlators of Generalized Twist Operators} we reviewed the results of \cite{Belin:2013uta} on various formal expressions regarding the conformal dimension $h_q(\mu,\lambda)$ and the magnetic flux response $k_q(\mu,\lambda)$ of the generalized twist operators. In this subsection, we use this formulae as well as the results of subsection \ref{subsec:EGBM} to calculate $h_q(\mu,\lambda)$ and $k_q(\mu,\lambda)$ in the boundary CFT dual to Einstein-Gauss-Bonnet-Maxwell theory \eqref{eq:einstein_gauss_bonnet_maxwell_action}. First we recall equation \eqref{eq:h_q in terms of energy den}, which we will reexpress in terms of $x_q$
\be\label{eq:h_q in terms of energy den recall}
h_q(\mu,\lambda)=2\pi\tilde{L}\frac{R^{d-1}}{d-1}q\(\mathcal{E}(1;0,\lambda)-\mathcal{E}(x_q;\mu,\lambda)\),
\ee
on using the fact the $T_0^{-1}=2\pi\tilde{L}$. Note also that by equations \eqref{eq:energy den}, \eqref{eq:m} and \eqref{eq:black_hole_energy}, we have that
\be
\mathcal{E}(1;0,\lambda)=\frac{m(0;1,\lambda)}{2\ell^{d-1}_P R^{d-1}}=\frac{d-1}{2 R^{d-1}}\frac{\tilde{L}^{d-2}}{\ell^{d-1}_P}\(\frac{1}{f_{\infty}}-1+\lambda f_{\infty}\) = 0.
\ee
Thus, combining the above, we take
\be\label{eq:h_q final}
h_q(\mu,\lambda) = -\pi \(\frac{\tilde{L}}{\ell_P} \)^{d-1} q \( \frac{1}{f_\infty}x_q^d-\( 1-\frac{d-2}{2(d-1)}\( \frac{\mu\ell_*}{2\pi \tilde{L}} \)^2 \)x_q^{d-2} +\lambda f_\infty x_q^{d-4} \).
\ee

One can verify that $h_q(0,0)$ gives the Einstein gravity result (remembering that $\tilde{L}=L$) and $h_q(0,\lambda)$ gives the Einstein-Gauss-Bonnet result in agreement with \cite{Hung:2011nu}, while $h_q(\mu,0)$ gives the Einstein-Maxwell result (remembering that $\tilde{L}=L$) also in agreement with \cite{Belin:2013uta}.

The expansion coefficients from \eqref{eq:d_q h_q Belin} are
\be\label{eq: h10 and h02}
h_{10}=\frac{2}{d-1}\pi^{1-\frac{d}{2}}\Gamma\(\frac{d}{2}\)\tilde{C}_T, \ \ \mathrm{and} \ \  h_{02}=-\frac{(d-2)(2d-3)}{4\pi(d-1)^2}\( \frac{\tilde{L}}{\ell_P} \)^{d-1}\(\frac{\ell_*}{\tilde{L}}\)^2.
\ee
These are in agreement with \cite{Hung:2011nu,Belin:2013uta}. Note, however, that one recovers $h_{02}$ from \cite{Belin:2013uta} by simply setting $L=\tilde{L}$.

To find the magnetic response, using equations \eqref{eq:mu in terms of q} and \eqref{eq:black_hole_charge} one can express the charge density as
\be\label{eq:Q charge den def}
\mathcal{Q}(x;\mu,\lambda)=\frac{d-2}{R^{d-1}}\(\frac{\tilde{L}}{\ell_P}\)^{d-1}\(\frac{\ell_*}{\tilde{L}}\)^2\frac{\mu}{4\pi}x^{d-2}.
\ee
Thus, the magnetic response is given by
\be\label{eq:k_q}
k_q(\mu,\lambda)=2\pi q R^{d-1}\mathcal{Q}(x_q;\mu,\lambda)=q (d-2)\(\frac{\tilde{L}}{\ell_P}\)^{d-1}\(\frac{\ell_*}{\tilde{L}}\)^2 \frac{\mu}{2}x_q^{d-2}.
\ee
This is trivially in agreement with \cite{Belin:2013uta} in the limit $\lambda\to 1$, since then $L=\tilde{L}$. Note also that $k_q(\mu,\lambda)$ depends implicitly on $\lambda$ through $x_q$. The expansion coefficients from \eqref{eq:d_q h_q Belin} are
\be\label{eq:k01 and k11}
k_{01}=\frac{d-2}{2}\( \frac{\tilde{L}}{\ell_P} \)^{d-1}\(\frac{\ell_*}{\tilde{L}}\)^2=(d-1)k_{11},
\ee
which are in agreement with \cite{Belin:2013uta} for $L=\tilde{L}$.

%%%-----------------------------------------------------------------------------------------------------------------------------------------------------------------
\subsection{Density Matrix Spectrum}
\label{subsec:spectrum}

As discussed in section \ref{sec:renyi}, \ren entropies can provide more information than entanglement entropy and more specifically, one can recover the whole spectrum of the density matrix $\rho_\mt{A}$. We have already seen in subsection \ref{subsec:renyi_basics} that the Min-Entropy limit of \ren entropy can provide the value of the largest eigenvalue of the density matrix spectrum, as shown in equation \eqref{eq:min_entropy}.

One can get even more information if they perform an $\frac{1}{q}$ expansion in the \ren entropy. For example, if we assume that the largest eigenvalue is equal to $\lambda_1$ and has degeneracy $d_1$, while the second largest eigenvalue is equal to $\lambda_2$ and has degeneracy $d_2$, such an expansion should look like
\begin{equation}
\begin{split}
{S_q} &= \frac{1}{{1 - q}}\ln {\mathop{\rm Tr}\nolimits} {\rho _\mt{A}}^q\\
 &= \frac{1}{{1 - q}}\left( {{d_1}{\lambda _1}^q + {d_2}{\lambda _2}^q +  \ldots } \right)\\
 &\simeq \frac{q}{{1 - q}}\left[ {\ln {\lambda _1} + \frac{1}{q}\ln {d_1} + \ln \left( {1 + \frac{{{d_2}}}{{{d_1}}}{{\left( {\frac{{{\lambda _2}}}{{{\lambda _1}}}} \right)}^q}} \right)} \right].
\end{split}
\label{eq:renyi_expansion_spectrum}
\end{equation}
Thus, an appropriate $\frac{1}{q}$ expansion of \ren entropy can provide the largest eigenvalue, its degeneracy, as well as the other eigenvalues, which appear is terms that are non-analytic in $\frac{1}{q}$. As such non-analytic terms do not appear from holographic calculations, the authors of \cite{Hung:2011nu} make several arguments that lead to the conjecture that the thermodynamic spectrum contains only one discrete eigenvalue and the rest is comprised of a continuum of states, with the spectral function of this continuum being analytical at the position of the delta function relative to the discrete eigenvalue. In the following, we will just calculate the dependence of the largest eigenvalue and its degeneracy on the higher derivative coupling (and thus, on the boundary theory central charges) and the chemical potential.

For this purpose, we find an $\frac{1}{q}$ expansion for the solution of equation \eqref{eq:xn_equation}. It can be shown that the former looks like
\begin{equation}
x_q = x_\infty + \frac{{{x_{\frac{1}{q}}}}}{q} + \mathcal{O}\left( {\frac{1}{{{q^2}}}} \right) ,
\end{equation}
where $x_\infty $ is given by equation \eqref{eq:xinfty} and
\begin{equation}
{x_{\frac{1}{q}}} = \frac{{x_\infty^2 - 2\lambda {f_\infty }}}{{\frac{{2dx_\infty^2}}{{{f_\infty }}} - \left( {d - 2} \right)\left( {1 + \frac{{d - 2}}{{2\left( {d - 1} \right)}}{{\left( {\frac{{\mu \ell_*}}{{2\pi \tilde{L}}}} \right)}^2}} \right)}} .
\end{equation}
Substituting in \eqref{eq:holographic_Renyi_entropy_result_simplified}, it is not difficult to show that
\begin{equation}
{S_q}\left( {\mu ,\lambda } \right) = \frac{{2\pi q}}{{q - 1}}{V_{\mathds{H}^{d-1}} }{\left( {\frac{{\tilde L}}{{{\ell_P}}}} \right)^{d - 1}}\frac{{d - 1}}{{d - 2}} \left[{S_\infty } + \frac{{{S_{\frac{1}{q}}}}}{q} + \mathcal{O}\left( {\frac{1}{{{q^2}}}} \right) \right],
\label{eq:renyi_large_q_expansion}
\end{equation}
where
\begin{align}
S_\infty &= {\frac{{x_1^d - x_\infty^d}}{{{f_\infty }}} - \lambda {f_\infty }\left( {x_1^{d - 4} - x_\infty^{d - 4}} \right) - \frac{{x_1^{d - 1}}}{{d - 1}} - 2\lambda {f_\infty }\frac{{x_1^{d - 3}}}{{d - 3}}} ,\\
{S_{\frac{1}{q}}} &=  - \frac{{dx_\infty^{d - 1}{x_{\frac{1}{q}}}}}{{{f_\infty }}} + \lambda {f_\infty }\left( {d - 4} \right)x_\infty^{d - 5}{x_{\frac{1}{q}}} + \frac{{x_\infty^{d - 1}}}{{d - 1}} + 2\lambda {f_\infty }\frac{{x_\infty^{d - 3}}}{{d - 3}} .
\end{align}

Comparing equations \eqref{eq:renyi_expansion_spectrum} and \eqref{eq:renyi_large_q_expansion}, the greatest eigenvalue and its degeneracy are given by
\begin{align}
{\lambda _1} &= \exp \left[ -{2\pi {V_{\mathds{H}^{d-1}}}{{\left( {\frac{{\tilde L}}{{{\ell _P}}}} \right)}^{d - 1}}\frac{{d - 1}}{{d - 2}}{S_\infty }} \right] ,\\
{d_1} &= \exp \left[ -{2\pi {V_{\mathds{H}^{d-1}}}{{\left( {\frac{{\tilde L}}{{{\ell _P}}}} \right)}^{d - 1}}\frac{{d - 1}}{{d - 2}}{S_{\frac{1}{q}}}} \right] .
\end{align}

%%%-----------------------------------------------------------------------------------------------------------------------------------------------------------------
\section{Discussion}
\label{sec:discussion}

Belin et al. \cite{Belin:2013uta} defined a new class of entropic entanglement measures which extend the definition of Entanglement \ren Entropy (ERE) to include a chemical potential for a conserved global charge in a CFT. We extended the holographic calculation of charged Entanglement \ren Entropy for a spherical entangling surface in theories that have a holographic dual with higher derivative Gauss-Bonnet terms. Using appropriate conformal transformations, as in previous calculations \cite{Hung:2011nu,Belin:2013uta}, the entanglement \ren entropy for a spherical entangling surface in Minkowski space can be calculated as the thermal entropy in the hyperbolic cylinder. The latter can be connected with the thermal entropy of asymptotically AdS topological black holes with hyperbolic horizons via the AdS/CFT dictionary.

A very interesting outcome of the holographic calculations of the \ren entropies is the violation of an inequality they must obey by definition. This is observed in theories that are characterised by sufficiently large Gauss-Bonnet coupling, not high enough though to violate causality. The definition \eqref{eq:Renyi_entropy_definition} of \ren entropies on a probability distribution implies that it always obeys $\frac{\partial }{{\partial q}}\left( {\frac{{q - 1}}{q}{S_q}} \right) \ge 0$. However, as shown in subsection \ref{subsec:ren entropy inequalities}, when \ren entropy is calculated holographically, the validity of the aforementioned inequality is intertwined with the positivity of the thermal entropy of the relative topological black holes. As higher derivative Gauss-Bonnet corrections are considered, the thermal entropy cannot be calculated on the basis of the Bekenstein-Hawking formula, but rather on the basis of Wald's formula. This, however, does not return necessarily positive results when the Gauss-Bonnet coupling is larger than $\lambda  > \frac{{\left( {d - 3} \right)\left( {{d^2} + d - 8} \right)}}{{4d{{\left( {d - 1} \right)}^2}}}$. Such high Gauss-Bonnet couplings are allowed by causality, thus we conclude that gravitational theories with higher derivative corrections, which correspond to a boundary CFT at a stable thermal ensemble, should be restricted by a constraint stricter than causality.\footnote{This \ren inequality violation is also evident in the results of \cite{Hung:2011nu}, although not mentioned by the authors.}

In \cite{Cvetic:2001bk} another possible resolution is conjectured, as a solution to the negative entropy problem. The authors claim that the exclusion of the corresponding parameter values as non-physical ones, may not be the appropriate solution to the problem, as the definition of this parameter region is dependent on the background space under consideration. It is true that for the parameters that the asymptotically AdS black hole has negative entropy, an asymptotically dS black hole has positive entropy. So, it is conjectured that some kind a phase transition between the asymptotically AdS and asymptotically dS black hole occurs. If this conjecture is true, then this region of parameters may contain information about the dS/CFT correspondence.

Independently of a deeper possible meaning of the negative entropy black holes and the violation of the \ren entropy inequality, we have to notice that they are both an artifact of higher derivative terms competing with the leading Einstein term. Thus, it is possible that in fully realistic M-theory or string theory compactification, where even higher order terms have to be taken into account, negative entropy black holes and \ren entropy inequality violation never occur.

Another interesting finding, consistent with the outcomes of \cite{Hung:2011nu}, is the fact that \ren entropies holographically calculated in higher derivative theories appear to be practically a linear function of the central charge ratio $\frac{\tilde{C}_T}{a_d^*}$ in the regime consistent with causality, although they are described by quite complicated expressions. If one considers a chemical potential, the dependence of the \ren entropy on $\frac{\tilde{C}_T}{a_d^*}$ becomes less linear. However, the linear dependence of \ren entropy on the central charge ratio $\frac{\tilde{C}_T}{a_d^*}$ is still a good approximation as long as $\frac{\mu\ell_*}{2\pi \tilde{L}} < 1$. Furthermore, there is a difference with the results of \cite{Hung:2011nu}. It turns out, that unlike the vanishing chemical potential case, where \ren entropy is a decreasing function of the Gauss-Bonnet coupling, above a critical chemical potential it becomes an increasing function of the latter. This critical chemical potential is an increasing function of $q$ with a finite limit as $q \to \infty$.

\ren entropies appear to be an increasing function of the chemical potential, which is also shown in \cite{Belin:2013uta}. This is valid also when Gauss-Bonnet corrections are added in the bulk theory. It turns out that the dependence of \ren entropy on the chemical potential is stronger for larger Gauss-Bonnet couplings. For large chemical potentials, \ren entropy is completely determined by the product of $\sqrt{f_\infty}$ with the chemical potential and appears to be independent of other factors, such as $q$.

The study of \ren entropies for imaginary chemical potential is also revealing some interesting features. Although no Hawking-Page phase transitions occur for asymptotically AdS black holes with hyperbolic horizons, which could be related to the confined and deconfined phases of the boundary CFT, there is a significant difference in the dependence of \ren entropy on the imaginary chemical potential between cases with $\lambda \geq 0$ and $\lambda<0$. Specifically, the analytic continuation of \ren entropies halts at a finite value of the imaginary chemical potential when $\lambda \geq 0$, unlike the case $\lambda<0$ that holds for arbitrarily high imaginary chemical potential. In the case of negative Gauss-Bonnet coupling there is also a significant difference between the $d = 4$ case and higher dimensions. Specifically, in higher dimensions, all \ren entropies have the same asymptotic behavior for large imaginary chemical potential, while for $d = 4$, the ratio $\frac{\tilde{S}_q(\mu_\mt{E})}{\tilde{S}_1(0)}$ tends to the finite value $\frac{q+1}{2q}$.

We would like also to point out that the Entanglement entropy acquired as the $q \to 1$ limit of \ren entropy recovers the thermal black hole entropy, being consistent with the Ryu-Takayanagi formula.

Two other interesting quantities are the conformal dimension $h_q(\mu,\lambda)$ and the magnetic flux response $k_q(\mu,\lambda)$ of the generalized twist operators. These quantities obey an interesting universal property which is apparent in our case as well. That is, the expressions simplify tremendously when one calculates their derivative with respect to $q$ evaluated at $q=1$ and $\mu=0$. This is evident from the expansion coefficients $h_{10},~h_{02}$ and $k_{01},~k_{11}$ of the conformal dimension and the magnetic flux response respectively. Interestingly, this universal property is the same in the case of \cite{Hung:2011nu}, as well as \cite{Belin:2013uta} for $\tilde{L}=L$.

A natural generalisation of the calculations presented here would be the inclusion of higher derivative gauge interactions, using the results of \cite{Anninos:2008sj}. It may also be of interest to extend these results in the case of the canonical ensemble, where the charge instead of the chemical potential is kept fixed.
%%%-----------------------------------------------------------------------------------------------------------------------------------------------------------------

\acknowledgments

We would like to thank Nils Carqueville for useful comments on a draft of this paper. The research of G.P. is supported and implemented under the ARISTEIA II action of the operational programme for education and long life learning and is co-funded by the European Union (European Social Fund) and National Resources of Greece.

%%%-----------------------------------------------------------------------------------------------------------------------------------------------------------------
\appendix

\section{Expansive Expressions for \ren Entropy}
\label{sec:expansion}

Equation \eqref{eq:xn_equation} can provide an expansion of \ren entropies around the vanishing chemical potential and Gauss-Bonnet limit, which is exactly solvable \cite{Hung:2011nu}. It is not difficult to acquire an expansive solution of \eqref{eq:xn_equation},
\begin{multline}
x_q = \frac{{1 + \sqrt {1 + d\left( {d - 2} \right){q^2}} }}{{dq}} + \frac{{{{\left( {d - 2} \right)}^2}q}}{{2\left( {d - 1} \right)\sqrt {1 + d\left( {d - 2} \right){q^2}} }}{\left( {\frac{{\mu {\ell _*}}}{{2\pi \tilde{L}}}} \right)^2}\\
 + 2\frac{{{{\left( {d - 1 - \sqrt {1 + d\left( {d - 2} \right){q^2}} } \right)}^2}}}{{d\left( {d - 2} \right)q\sqrt {1 + d\left( {d - 2} \right){q^2}} }}\lambda \\
 + \frac{{d{{\left( {d - 1} \right)}^2}{{\left( {d - 2} \right)}^2}{q^4} - 2\left( {1 + d\left( {d - 2} \right){q^2}} \right){{\left( {1 - \sqrt {1 + d\left( {d - 2} \right){q^2}} } \right)}^2}}}{{2\left( {d - 1} \right)\left( {d - 2} \right)q{{\left( {\sqrt {1 + d\left( {d - 2} \right){q^2}} } \right)}^3}}}\lambda {\left( {\frac{{\mu {\ell _*}}}{{2\pi \tilde{L}}}} \right)^2} \\ + \mathcal{O}\(\mu^2,\lambda^2\) .
\end{multline}
Based on the above, substituting $x_q$ at \eqref{eq:holographic_Renyi_entropy_result_simplified} and \eqref{eq:holographic _entaglement_entropy_result}, and expanding the Gauss-Bonnet coupling in terms of
$\frac{\tilde{C}_T}{a_d^*} $, it is a matter of algebra to find that, for example for $d=4$ it is true that
\begin{multline}
\frac{{{S_q}\left( {\mu ;4} \right)}}{{{S_1}\left( {0;4} \right)}} = \frac{1}{{64\left( {q - 1} \right){q^3}}}\left\{ { - 1 - 12{q^2} + 40{q^4} - {{\left( {1 + 8{q^2}} \right)}^{\frac{3}{2}}}} \phantom{\left( {\frac{{{\tilde{C}_T}}}{{{a_d}^*}} - 1} \right)} \right.\\
 + 4{q^2}\left( { - 1 + 4{q^4} - {{\left( {1 + 8{q^2}} \right)}^{\frac{1}{2}}}} \right){\left( {\frac{{\mu {\ell _*}}}{{2\pi \tilde{L}}}} \right)^2}
\left[ { - \frac{9}{4}\left( {(1 + 8{q^4} + \left( {1 - 4{q^2}} \right){{\left( {1 + 8{q^2}} \right)}^{\frac{1}{2}}}} \right)} \phantom{{\left( {\frac{{\mu {\ell _*}}}{{2\pi \tilde{L}}}} \right)}^2} \right.\\
 + \left. {\left. {4{q^2}\left( { - 2 + 5{q^2} - \frac{{2 + 7{q^2}}}{{{{\left( {1 + 8{q^2}} \right)}^{\frac{1}{2}}}}}} \right){{\left( {\frac{{\mu {\ell _*}}}{{2\pi \tilde{L}}}} \right)}^2}} \right]\left( {\frac{{{\tilde{C}_T}}}{{{a_d}^*}} - 1} \right)} \right\} + \mathcal{O}\left( {{\left( \frac{\tilde{C}_T}{a_d^*} \right)^2},{\mu ^4}} \right) .
\end{multline}

From the expansion of $\frac{{{S_q}\left( {\mu ;4} \right)}}{{{S_1}\left( {0;4} \right)}}$, it is evident that the linear term in $\frac{\tilde{C}_T}{a_d^*}$ changes sign when the chemical potential equals
\begin{equation}
{\left( {\frac{{{\mu _{\mathrm{cr}}}\left( {q;4} \right){\ell _*}}}{{2\pi \tilde{L}}}} \right)^2} = \frac{{9\left( {(1 + 8{q^4} + \left( {1 - 4{q^2}} \right){{\left( {1 + 8{q^2}} \right)}^{\frac{1}{2}}}} \right)}}{{16{q^2}\left( { - 2 + 5{q^2} - \frac{{2 + 7{q^2}}}{{{{\left( {1 + 8{q^2}} \right)}^{\frac{1}{2}}}}}} \right)}} ,
\end{equation}
which is an increasing function of $q$, as explained in subsection \ref{subsec:Renyi}.

From the expansion of $\frac{{{S_q}\left( {\mu ;4} \right)}}{{{S_1}\left( {\mu;4} \right)}}$, one can check that the linear term in $\frac{\tilde{C}_T}{a_d^*}$ is affected very weakly by the chemical potential, as discussed in subsection \ref{subsec:Renyi}.
\begin{multline}
\frac{{{S_q}\left( {\mu ;4} \right)}}{{{S_1}\left( {\mu ;4} \right)}} = \frac{1}{{64\left( {q - 1} \right){q^3}}}\left\{ { - 1 - 12{q^2} + 40{q^4} - {{\left( {1 + 8{q^2}} \right)}^{\frac{3}{2}}}} \phantom{\left( \frac{\tilde{C}_T}{a_d^*}-1 \right)} \right.\\
 + \frac{1}{3}\left( {1 + 8{q^4} + \left( {1 - 4{q^2}} \right){{\left( {1 + 8{q^2}} \right)}^{\frac{1}{2}}}} \right){\left( {\frac{{\mu {\ell _*}}}{{2\pi \tilde L}}} \right)^2}
\left[ { - \frac{9}{4}\left( {(1 + 8{q^4} + \left( {1 - 4{q^2}} \right){{\left( {1 + 8{q^2}} \right)}^{\frac{1}{2}}}} \right)} \phantom{\left( {\frac{{\mu {\ell _*}}}{{2\pi L}}} \right)^2} \right.\\
 + \left. {\left. {\frac{1}{4}\left( {5 - 8{q^2} + 24{q^4} + \frac{{5 + 12{q^2} - 80{q^4}}}{{{{\left( {1 + 8{q^2}} \right)}^{\frac{1}{2}}}}}} \right){{\left( {\frac{{\mu {\ell _*}}}{{2\pi L}}} \right)}^2}} \right]\left( {\frac{{{\tilde{C}_T}}}{{{a_d}^*}} - 1} \right)} \right\} + \mathcal{O}\left( {{\left( \frac{\tilde{C}_T}{a_d^*} \right)^2},{\mu ^4}} \right)
\end{multline}

For larger dimensions, we give only the expansion for $\frac{S_\infty \left( \mu \right)}{S_1 \left( 0 \right)}$,
\begin{multline}
\frac{{{S_\infty }\left( {\mu ;d} \right)}}{{{S_1}\left( {0;d} \right)}} = \left( {\frac{{d - 2}}{{d - 1}} - {{\left( {\frac{{d - 2}}{d}} \right)}^{\frac{d}{2}}}} \right) + \frac{{{{\left( {d - 2} \right)}^2}}}{{4\left( {d - 1} \right)}}\left( {1 - {{\left( {\frac{{d - 2}}{d}} \right)}^{\frac{{d - 2}}{2}}}} \right){\left( {\frac{{\mu {\ell _*}}}{{2\pi \tilde{L}}}} \right)^2}\\
 + \left[ { - \frac{{\left( {d - 1} \right){{\left( {\frac{{d - 2}}{d}} \right)}^{\frac{d}{2}}}}}{{2\left( {d - 2} \right)}} + \frac{{{{\left( {d - 2} \right)}^2}}}{8}\left( {1 - {{\left( {\frac{{d - 2}}{d}} \right)}^{\frac{{d - 2}}{2}}}} \right){{\left( {\frac{{\mu {\ell _*}}}{{2\pi \tilde{L}}}} \right)}^2}} \right]\left( {\frac{{{\tilde{C}_T}}}{{{a_d}^*}} - 1} \right) \\
 + \mathcal{O}\left( {{\left( \frac{\tilde{C}_T}{a_d^*} \right)^2},{\mu ^4}} \right) .
\end{multline}
The above results in the conclusion that the critical value of the chemical potential, above which all $\frac{S_q \left( \mu \right)}{S_1 \left( 0 \right)}$ become an increasing function of $\frac{\tilde{C}_T}{a_d^*}$ is a decreasing function of the number of dimensions
\begin{equation}
{\left( {\frac{{{\mu _{\mathrm{cr}}}\left( {\infty ;d} \right){\ell _*}}}{{2\pi \tilde{L}}}} \right)^2} = \frac{{4\left( {d - 1} \right)}}{{d{{\left( {d - 2} \right)}^2}\left( {{{\left( {\frac{{d - 2}}{d}} \right)}^{ - \frac{{d - 2}}{2}}} - 1} \right)}} ,
\end{equation}
as explained in subsection \ref{subsec:Renyi}.

In subsection \ref{subsec:Renyi}, it is also claimed that the ratios $\frac{S_q \left( \mu \right)}{S_1 \left( 0 \right)}$ and $\frac{S_q \left( \mu \right)}{S_1 \left( \mu \right)}$ are almost linear functions of in the ratio of the central charge ration $\frac{{{C_T}}}{{{a_d}^*}}$ in the regime of Gauss-Bonnet couplings allowed by causality. In the expansion we just performed, this can be shown, keeping one more term. Because of the complexity of the formulas, we show the result only for the $q \to \infty$ limit. We find
\begin{multline}
\frac{{{S_\infty }\left( {\mu ;4} \right)}}{{{S_1}\left( {0;4} \right)}} = \frac{1}{8}\left[ {5 + 2{{\left( {\frac{{\mu {\ell _*}}}{{2\pi \tilde L}}} \right)}^2}} \right] + \frac{1}{{32}}\left[ { - 9 + 10{{\left( {\frac{{\mu {\ell _*}}}{{2\pi \tilde L}}} \right)}^2}} \right]\left( {\frac{{{C_T}}}{{{a_d}^*}} - 1} \right)\\
 + \frac{1}{{128}}\left[ { - 3 - 2{{\left( {\frac{{\mu {\ell _*}}}{{2\pi \tilde L}}} \right)}^2}} \right]{\left( {\frac{{{C_T}}}{{{a_d}^*}} - 1} \right)^2} + \mathcal{O}\left( {{\left( \frac{\tilde{C}_T}{a_d^*} \right)^3},{\mu ^4}} \right)
\end{multline}
and
\begin{multline}
\frac{{{S_\infty }\left( {\mu ;4} \right)}}{{{S_1}\left( {\mu ;4} \right)}} = \frac{1}{8}\left[ {5 + \frac{1}{3}{{\left( {\frac{{\mu {\ell _*}}}{{2\pi \tilde L}}} \right)}^2}} \right] + \frac{1}{{32}}\left[ { - 9 + 3{{\left( {\frac{{\mu {\ell _*}}}{{2\pi \tilde L}}} \right)}^2}} \right]\left( {\frac{{{C_T}}}{{{a_d}^*}} - 1} \right)\\
 + \frac{1}{{128}}\left[ { - 3 + 17{{\left( {\frac{{\mu {\ell _*}}}{{2\pi \tilde L}}} \right)}^2}} \right]{\left( {\frac{{{C_T}}}{{{a_d}^*}} - 1} \right)^2} + \mathcal{O}\left( {{\left( \frac{\tilde{C}_T}{a_d^*} \right)^3},{\mu ^4}} \right).
\end{multline}
It is also evident that the inclusion of a chemical potential make the linear approximation of the above expressions worse, especially for the ratio $\frac{S_q \left( \mu \right)}{S_1 \left( \mu \right)}$, as also commented in subsection \ref{subsec:Renyi}.

%%%-----------------------------------------------------------------------------------------------------------------------------------------------------------------
\small

\newcommand\arxiv[2]      {\href{http://arXiv.org/abs/#1}{#2}}
\newcommand\doi[2]        {\href{http://dx.doi.org/#1}{#2}}
\newcommand\httpurl[2]    {\href{http://#1}{#2}}

\end{document}